\documentclass[times,twocolumn,final]{elsarticle}

\let\today\relax
\makeatletter
\def\ps@pprintTitle{%
    \let\@oddhead\@empty
    \let\@evenhead\@empty
    \def\@oddfoot{\footnotesize\itshape
         {} \hfill\today}%
    \let\@evenfoot\@oddfoot
    }
\makeatother

\usepackage{xcolor}
\usepackage{svg}

\usepackage{framed,multirow}
\usepackage{siunitx}
\usepackage{epsfig,subfig}
\newcommand{\expon}[1]{\num[round-mode=places, round-precision=1]{#1}}
\newif\ifshowtop
\showtopfalse

\usepackage{amssymb,amsmath}
\usepackage{latexsym}
\usepackage{booktabs}
\usepackage{mathtools}   
\usepackage{url}
\usepackage{xcolor}
\definecolor{newcolor}{rgb}{.8,.349,.1}
\usepackage{hyperref}
\usepackage[switch,pagewise]{lineno} 
\begin{document}
\begin{frontmatter}
\title{
FlowMotion: Target-Predictive Conditional Flow Matching for Jitter-Reduced Text-Driven Human Motion Generation}

\author[1]{Manolo Canales Cuba\corref{cor1}}
\author[1]{Vinícius do Carmo Melício} 
\cortext[cor1]{Corresponding author: manolo.canales@ufabc.edu.br}
\author[1]{João Paulo Gois} 
\address[1]{Universidade Federal do ABC, Santo André, Brazil}

\begin{abstract}
Achieving high-fidelity and temporally smooth 3D human motion generation remains a challenge, particularly within resource-constrained environments. We introduce FlowMotion, a novel method leveraging Conditional Flow Matching (CFM). FlowMotion incorporates a training objective within CFM that focuses on more accurately predicting target motion in 3D human motion generation, resulting in enhanced generation fidelity and temporal smoothness while maintaining the fast synthesis times characteristic of flow-matching-based methods. FlowMotion achieves state-of-the-art jitter performance, achieving the best jitter in the KIT dataset and the second-best jitter in the HumanML3D dataset, and a competitive FID value in both datasets. This combination provides robust and natural motion sequences, offering a promising equilibrium between generation quality and temporal naturalness.
\end{abstract}

\end{frontmatter}

\section{Introduction} 

The synthesis of 3D human body motion has diverse applications across fields such as robotics \cite{jiang2024harmon,cheng2024expressive,annabi2024unsupervised}, VR/AR \cite{du2023avatars,luo2024real}, entertainment \cite{yang2024smgdiff, peng2022ase,starke2020local,Starke_2019},
and social interaction in virtual 3D spaces 
\cite{jiang2024autonomous, zhang2020generating,hassan2021populating, hassan2021stochastic}.  While recent advances in 3D human motion generation are significant, several challenges remain. The inherent complexities of motion generation are exacerbated by the need to incorporate diverse constraints, such as spatial trajectories~\cite{barquero2024seamless,cohan2024flexible}, interactions with surrounding objects~\cite{huang2023diffusion, yi2024tesmo}, or temporal specifications defined by keyframes~\cite{karunratanakul2023guided}, all aimed at producing lifelike movements.

To achieve realistic and context-aware motion synthesis, motion generation techniques frequently leverage data-driven motion capture data. These techniques include, for instance, approaches based on deep learning~\cite{actorvae,Tevet_motionclip2022}, reinforcement learning~\cite{peng2017deeploco,Peng_2018}, and hybrid methods~\cite{yuan2023physdiff}, which utilize these datasets to model and predict complex movements.

Recent studies in generative models have driven the development of new techniques for synthesizing 3D human body motion. These methods primarily focus on generating realistic movements based on user inputs, especially descriptive text that specifies the intended action. A key advantage of these generative approaches is their ability to produce a diverse range of plausible motion sequences from a single prompt. This allows users to explore multiple interpretations of a desired movement and select the sequence that best aligns with their creative vision.  Among these generative methods, classes such as variational autoencoders (VAEs)~\cite{zhang2023generating,Petrovich2022,Guo_2022}, generative adversarial networks (GANs)~\cite{hernandez2019human,degardin2022generative,barsoum2018hp,ahn2018text2action}, and diffusion models~\cite{tevet2022human,MotionDiffuse,dabral2023mofusion} have received great attention for 3D human motion generation, each offering distinct advantages and limitations. While VAEs have been used in motion generation, their output diversity is often limited due to the issue of posterior collapse. Similarly, GANs can suffer from mode collapse, making it challenging to achieve sufficient diversity in the generated motions. In contrast, diffusion models excel in generating diverse and natural motions, albeit with increased latency during the generation process.

Hybrid models, on the other hand, aim to minimize the limitations of individual architectures by combining their complementary strengths. Recent proposals include integrating features from GANs and diffusion models to enhance generation speed without sacrificing generalization capability~\cite{zhou2023emdm}. Furthermore, there are also approaches combining the strengths of GANs, VAEs, and diffusion models, utilizing distillation techniques for even faster sampling \cite{dai2024motionlcm}. Additionally, diffusion techniques have been applied to latent spaces to enable faster sampling, using VAEs \cite{mld}.

Beyond the aforementioned generative approaches, flow-matching-based models have demonstrated promising performance in generating both images~\cite{lipman2022flow, huang2024flow} and 3D human motion~\cite{hu2023motion}, achieving fast synthesis with adequate diversity in the results. While the motion synthesis approach by Hu et al.~\cite{hu2023motion} performs well when adequate computational resources (specifically GPU memory enabling larger batch sizes) are available, we observed challenges under hardware constraints. The necessity of using smaller batch sizes due to limited GPU memory can produce temporal inconsistencies, resulting in jitter effect in the generated motion. Even with careful tuning, achieving stable training with the standard conditional flow matching formulation presented difficulties under these conditions. We hypothesize that this artifact, potentially linked to the inherent variations in the conditional flow matching (CFM) formulation~\cite{lipman2022flow} itself (due to differences between complex data representations and the target Gaussian distribution), is  amplified by the training instability introduced when using smaller batch sizes necessitated by hardware limitations.

To harness the computational efficiency of flow matching while addressing the jitter effect, we introduce a training objective designed to approximate the original, unperturbed motion—effectively reducing fluctuations in the vector field during synthesis. This objective is implemented within the Conditional Flow Matching framework and focuses on directly predicting the target motion, leading to a more stable and accurate vector field. This approach is conceptually aligned with a recent formulation introduced by Lipman et al.~\cite{lipman2024flow}. However, our novel contribution lies in the application of this method to 3D human motion generation, specifically tailored to mitigate jitter.

To quantitatively evaluate the temporal smoothness of the generated motion, we employ an acceleration-based metric, which offers an objective and interpretable measure of motion stability and fidelity.

\begin{figure}[t!]
    \centering    \includegraphics[width=0.48\textwidth]{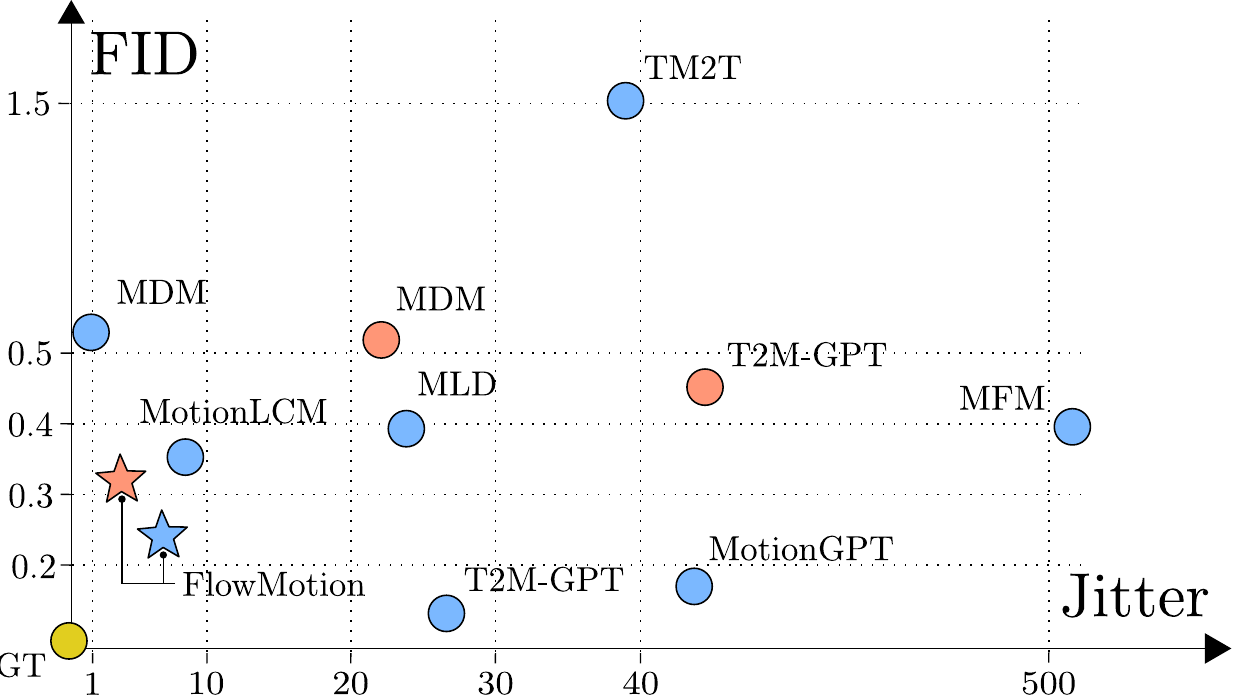} 
    \caption{FID-Jitter evaluation on the HumanML3D and KIT datasets. Jitter represents the absolute difference from the ground truth Jitter. HumanML3D is represented in light blue, KIT in orange. Stars of the same color denote our results: FlowMotion. FlowMotion proximity to the Ground Truth (GT), indicated in yellow at the origin, highlights its effectiveness in achieving a balance between generation fidelity, measured by Fréchet Inception Distance (FID), and motion smoothness, quantified by lower Jitter values, indicating a generation closer to real-world motion compared to existing methods.}
    \label{fig:jittercompare}
\end{figure} 

Our work introduces a novel application of a CFM training objective that directly predicts target motion to 3D human motion generation, offering enhanced generation fidelity and temporal smoothness. While maintaining a lightweight model architecture suitable for resource-constrained environments, our approach also retains the fast synthesis times characteristic of flow-matching-based methods. The application of this training objective to 3D human motion generation results in a method, which we call FlowMotion. Experimental results indicate that FlowMotion achieves a promising equilibrium between generation quality (FID) and temporal naturalness (Jitter) (Fig.~\ref{fig:jittercompare}). FlowMotion exhibits top-tier performance, achieving the best jitter in the KIT dataset and the second-best jitter in the HumanML3D dataset, while attaining a very competitive FID value in both datasets. This closely approximates real human motion, bolstering the ability to generate smooth, realistic animations. This enhanced fidelity and smoothness highlight the potential of our approach to generate robust and natural motion sequences.

This paper is organized as follows: In Sec.~\ref{secrelatedwork}, we present the related work on generative methods for 3D human motion synthesis. In Sec.~\ref{secmethod}, we present our method, covering both theoretical foundations and computational details. Sec.~\ref{secexperiments} describes the experiments conducted to evaluate the effectiveness of our approach. In Sec.~\ref{sec:results}, we discuss and analyze the results of these experiments. Finally, we conclude our study in Sec.~\ref{secconclusion}, highlighting potential directions for future work based on the insights gained from our approach.

\section{Related Work}\label{secrelatedwork}
We review techniques for 3D human motion generation in two subsections. Section~\ref{subsec:3d_motion_gen} provides an overview of generative models, ranging from GANs and VAEs to recent Transformer-based architectures, discussing their challenges and corresponding contributions. Section~\ref{subsec:diffusion_flow_matching} focuses on diffusion models, which have garnered significant attention and improvements, and flow matching models due to their shared characteristics.

\subsection{3D Human Motion Generation}
\label{subsec:3d_motion_gen}
Diverse studies have explored the synthesis of human body movements conditioned on text \cite{tevet2022human,Guo_2022,MotionDiffuse}. For instance, GANs have been used for generating movements based on specific actions described in text \cite{degardin2022generative}. Similarly, Text2Action~\cite{ahn2018text2action} employs a Seq2Seq-inspired model \cite{sutskever2014sequence} to translate text into motion, focusing on upper body movements. 

More recently, Guo et al.~\cite{Guo_2022} introduced a temporal VAE enhanced with attention mechanisms for generating human motion from text, along with the widely adopted HumanML3D dataset and associated evaluation metrics. This dataset, in conjunction with the KIT-ML dataset~\cite{kitml}, represents one of the largest currently available and has become a standard benchmark for subsequent research, including the present work.

Subsequent advancements have seen the incorporation of Transformer architectures. ACTOR~\cite{petrovich2021action} utilized a Transformer-based VAE, while TEACH~\cite{athanasiou2022teach} conditioned motion generation on sequences of textual motion descriptions using a similar architectural approach. Further, TM2T~\cite{guo2022tm2t} established connections between text tokens and motion sequences via Transformers, demonstrating efficacy on both HumanML3D and KIT-ML~\cite{kitml} datasets. T2M-GPT~\cite{zhang2023generating} employed a Vector Quantized Variational AutoEncoder (VQ-VAE) in conjunction with a CLIP text encoder~\cite{clip} to achieve a discrete representation of movement, exhibiting robust generalization capabilities.  Similarly, MotionGPT~\cite{jiang2023motiongpt}, based on the same VQ-VAE, constructs a vocabulary of movements combined with an advanced Transformer (T5~\cite{raffel2020exploring}) for integration with text. Text2Gestures~\cite{bhattacharya2021text2gestures} utilizes a Transformer encoder to process descriptive text encoded with GloVe~\cite{pennington2014glove} and a Transformer decoder which, combined with previous movements, generates subsequent ones, enabling the production of expressive animations.

\subsection{Diffusion and Flow Matching Models}
\label{subsec:diffusion_flow_matching}
Diffusion models, based on stochastic processes, have achieved significant success in tasks such as image \cite{dhariwal2021diffusion,ramesh2022hierarchical}, sound \cite{huang2023noise2music,zhu2023ernie}, and  video generation \cite{fei2024dysen}. Recently, several methods have leveraged diffusion models for motion generation~\cite{tevet2022human,MotionDiffuse}. Among these, MDM \cite{tevet2022human} employs a Transformer encoder to generate motion sequences through iterative denoising. The model optimizes a loss function that directly targets the final, denoised motion state. MDM integrates CLIP for text encoding and has been evaluated on standard benchmarks, supporting both text-conditioned and unconditional generation. Its architecture has been widely adopted, including in work by Cohan et al. \cite{cohan2024flexible}, which extends it to generate motions constrained by keyframes and trajectories.

Likewise, MotionDiffuse~\cite{MotionDiffuse} employs a more extensive network than MDM, using a Transformer encoder for noisy motion input and a Transformer decoder, specifically a cross-attention mechanism, for handling the input text. It uses CLIP for text encoding and is evaluated on established datasets. A similar approach is taken in FLAME \cite{kim2023flame}, which uses RoBERTa~\cite{liu2019roberta} for text encoding. Recently, attention networks combined with databases to retrieve movements that approximate textual input descriptions have been used alongside diffusion models to achieve better generalization in motion generation \cite{zhang2023remodiffuse}. MoFusion~\cite{dabral2023mofusion}, based on diffusion, does not use a Transformer for processing sequential motion data but instead employs a 1D-UNet, similar to that used in Stable Diffusion~\cite{rombach2022high} for images. It extends the generation process to be conditioned not only on text but also on audio.

However, these diffusion-based approaches tends to be slow in the inference process due to the extensive sequence of steps required for sampling. Efforts like MLD~\cite{mld} improve inference speed by performing diffusion in a latent space achieved through a prior VAE, reducing computational resources and accelerating inference. Similarly, MotionLCM~\cite{dai2024motionlcm} and EMDM~\cite{zhou2023emdm} achieve real-time inference by accelerating sampling, omitting denoising steps in the diffusion pipeline through distillation and a discriminator, respectively, in a manner analogous to techniques used in GANs. To ensure coherent body generation and prevent artifacts such as levitation, PhysDiff~\cite{yuan2023physdiff} integrate physical constraints during the sampling phase of the diffusion framework, though such constraints may reduce generation speed.

On the other hand, text-conditioned human motion generation via flow matching has shown promising results~\cite{hu2023motion}. Using the same architecture as MDM, this approach achieves better generalization and faster inference due to the streamlined trajectory in the sampling process. However, the use of flow matching relies on intractable integrals. Thus, in practice, one uses the Conditional Flow Matching (CFM) model~\cite{lipman2022flow}, which can generate erratic or jittering motions during inference, particularly under limited computational resources. The variations introduced by the CFM formulation, where the learned vector field is influenced by the interpolation between data representations and random Gaussian noise, may contribute to this artifact. In our method, we propose reducing jitter by modifying the loss function of the CFM, such that it directly compares the generated image with the noise-free original image. 

Distinct from previous work, our results indicate that the technique presented herein delivers visually superior results, exhibiting both higher generation fidelity and improved temporal smoothness, without increasing computational cost.

\section{The FlowMotion Method}\label{secmethod} 
Our method, FlowMotion, focuses on generating text-conditioned human motion sequences with high fidelity and temporal smoothness. To achieve this equilibrium between generation quality and naturalness, we leverage Conditional Flow Matching  (CFM) \cite{lipman2022flow}, which enables the learning of complex motion distributions.  In contrast to standard flow matching approaches, CFM operates on conditional distributions, which is advantageous for sampling as it promotes straighter trajectories and therefore, faster sampling.

Our training objective directly predicts the target motion, enhancing stability and reducing jitter compared to prior CFM
and diffusion-based approaches.  By employing a Transformer-based architecture, our method efficiently processes motion and text embeddings, enabling generating motion sequences exhibiting strong diversity with the provided textual descriptions.

In this section, we provide a concise overview of flow matching, followed by the details of our CFM-based framework, including the training procedure and sampling strategy.

\subsection{Flow Matching}\label{secflowmatching}
The flow matching~\cite{lipman2022flow} determines a time-dependent vector field $v: [0,1]\times \mathbb{R}^d \rightarrow \mathbb{R}^d$, that transforms a simpler probability density function $p_0$, such as a Gaussian distribution, into a more complex one, $p_1$, through the so-called {\it probability density path} $p:~[0,1]~\times~\mathbb{R}^d~\rightarrow~\mathbb{R}^+$, where $p_0 = p(0, \cdot)$ at the initial time $t=0$, and $p_1=p(1,\cdot)$ at the final time $t=1$.

The vector field $v$ defines the ordinary differential equation (ODE):
\begin{equation} \label{eq:ode}
\begin{aligned}
    \frac{d}{dt}\phi(t,x) & = v(t,\phi(t,x)), \\
    \phi(0,x) & = x,
\end{aligned}    
\end{equation}
where its solution $\phi: [0,1] \times \mathbb{R}^d \rightarrow \mathbb{R}^d$ is named \emph{flow,} 
the diffeomorphism induced by the vector field  $v$. This ensures that $\phi$ possesses a differentiable inverse, guaranteeing a smooth and invertible mapping of the probability space. For notational convenience, we use $\phi_t$ to denote $\phi(t,\cdot)$ and $v_t$ to denote $v(t,\cdot)$.

According to Chen et al.~\cite{chennn}, one can reparameterize the vector field $v$ using a neural network with parameters $\theta \in\mathbb{R}^\ell$. Consequently, the flow $\phi$ is also parameterized by $\theta$, resulting in a \textit{Continuous Normalizing Flow} (CNF). This CNF transforms the initial density $p_0$ to the density $p_t$ at time $t$ via a push-forward operation, specifically a change of variables:
\begin{equation*}
p_t = [\phi_t]_ * p_0.
\end{equation*}
This transformation is defined as:
\begin{equation*}
    p_t(y) = p_0(x) \left| \det \left( \frac{\partial \phi_t^{-1}(y)}{\partial y} \right) \right|,
\end{equation*}
where $x = \phi_t^{-1}(y)$ for $t\in [0,1]$.

Given a finite set of samples $x_1$ from an unknown data distribution $q$, and initializing with $p_0 = \mathcal{N}(0, I)$, Lipman et al.~\cite{lipman2022flow} introduce the concept of a \emph{conditional probability path} $p_t(\cdot|x_1) : \mathbb{R}^d \rightarrow \mathbb{R}^+$, defined for each sample $x_1$. At the final time $t=1$, $p_1(\cdot|x_1)$ is defined as a Gaussian distribution centered at $x_1$ with a small standard deviation $\sigma_{\min} > 0$, concentrating the probability around the sample. The distribution $p_t(x)$ is then defined as the marginalization of these conditional probability paths:
\begin{equation}\label{eq:marginalp}
    p_t(x) = \int p_t(x|x_1) q(x_1) dx_1.
\end{equation}

Thus, at $t=1$, the marginal distribution $p_1$ approximates the unknown distribution $q$. 

However, the direct computation of $p_t(x)$ via Eq.~\eqref{eq:marginalp} is intractable. Lipman et al.~\cite{lipman2022flow} demonstrate that an objective function designed to approximate the unknown vector field $v_t$, which generates the marginal probability path $p_t$, has gradients identical, with respect to the model parameters, to an objective function that approximates the conditional vector field $u_t(\cdot|x_1)$. This crucial property avoids the intractability of computing $p_t(x)$. Consequently, it suffices to define appropriate conditional probability paths $p_t(\cdot|x_1)$ and conditional vector fields $u_t(\cdot|x_1)$ to minimize the objective function with respect to these conditional vector fields.

Thus, for each $x_1 \sim q$, a conditional Gaussian probability path is defined for each time $t \in [0,1]$:
\begin{equation*}
    p_t(x|x_1) = \mathcal{N}(x; \mu_t(x_1), \sigma_t(x_1)^2I),
\end{equation*}
where, as defined previously, at time $t=1$: $p_1(x|x_1) = \mathcal{N}(x;x_1,\sigma_{\text{min}}^2I)$, with $\mu_1(x_1) =x_1$ and $\sigma_1(x_1)=\sigma_{\text{min}}$. Similarly, at $t=0$, $p_0(x|x_1) =\mathcal{N}(x;0,I)$, where $\mu_0(x_1)=0$ and $\sigma_0(x_1)=~1$, corresponding to the standard normal distribution. Finally, for $0<t<1$,  we define a conditional probability path where the mean and standard deviation are linearly interpolated between the boundary parameters $\{\mu_0, \sigma_0\}$ and $\{\mu_1, \sigma_1\}$:  
\begin{align*}
    \mu_t(x_1) &= tx_1, \\
    \sigma_t(x_1) &= 1 - (1 - \sigma_{\text{min}})t.
\end{align*}
This results in the general form:  
\begin{equation*}
    p_t(x|x_1) = \mathcal{N}\left(x; tx_1, \left(1 - (1 - \sigma_{\text{min}})t\right)^2I\right).
\end{equation*}

Under the Gaussian probability path $p_t$, an element from the range of the flow at time $t$ is given by $\psi_t(x) \in \mathbb{R}^d$, which is a function of the linear parameters $\mu_t(x_1)$ and $\sigma_t(x_1)$, as follows:
\begin{equation}\label{eq:psigauss_expanded}
    \psi_t(x) = \left(1 - (1-\sigma_{min})t\right)x + tx_1.
\end{equation}

The original \emph{Conditional Flow Matching} (CFM) objective~\cite{lipman2022flow} is formulated with respect to the initial sample $x_0$. However, for our iterative sampling process, we redefine the objective in terms of the intermediate state $x_t$. Utilizing the conditional vector field construction from Lipman et al.~\cite{lipman2022flow}, which expresses the vector field based on the means, standard deviations, and their derivatives. Thus, we adopt their formulation to define the conditional vector field:
\begin{equation}\label{eq:campotheorema}
    u_t(x_t|x_1) = \frac{x_1 - (1-\sigma_{min})x_t}{1-(1-\sigma_{min})t},
\end{equation}
where $x_t=\psi_t(x)$.

Rather than training a model to approximate $u_t$ directly,  we predict the target $x_1$ through a neural network conditioned on $x_t$ (Sec.~\ref{sec:training}). A similar formulation has also been recently proposed by Lipman~\cite{lipman2024flow} as an alternative parameterization of the vector field $u_t$. Given the predicted $x_1$ and the fixed $\sigma_{\text{min}}$, the vector field $u_t$ is computed via Eq.~\eqref{eq:campotheorema}. This approach
bypasses explicit vector field estimation while preserving the theoretical guarantees of CFM.

\subsection{Framework}
\label{sec:framework}
We start from the conditional probability path proposed by Lipman et al.~\cite{lipman2022flow}, as detailed previously, along with its associated flow and conditional vector field. These constructs are based on the interpolation of parameters of Gaussian distributions along a temporal trajectory defined by $t \in [0,1]$.

\begin{figure}[h!]
    \centering    \includegraphics[width=0.48\textwidth]{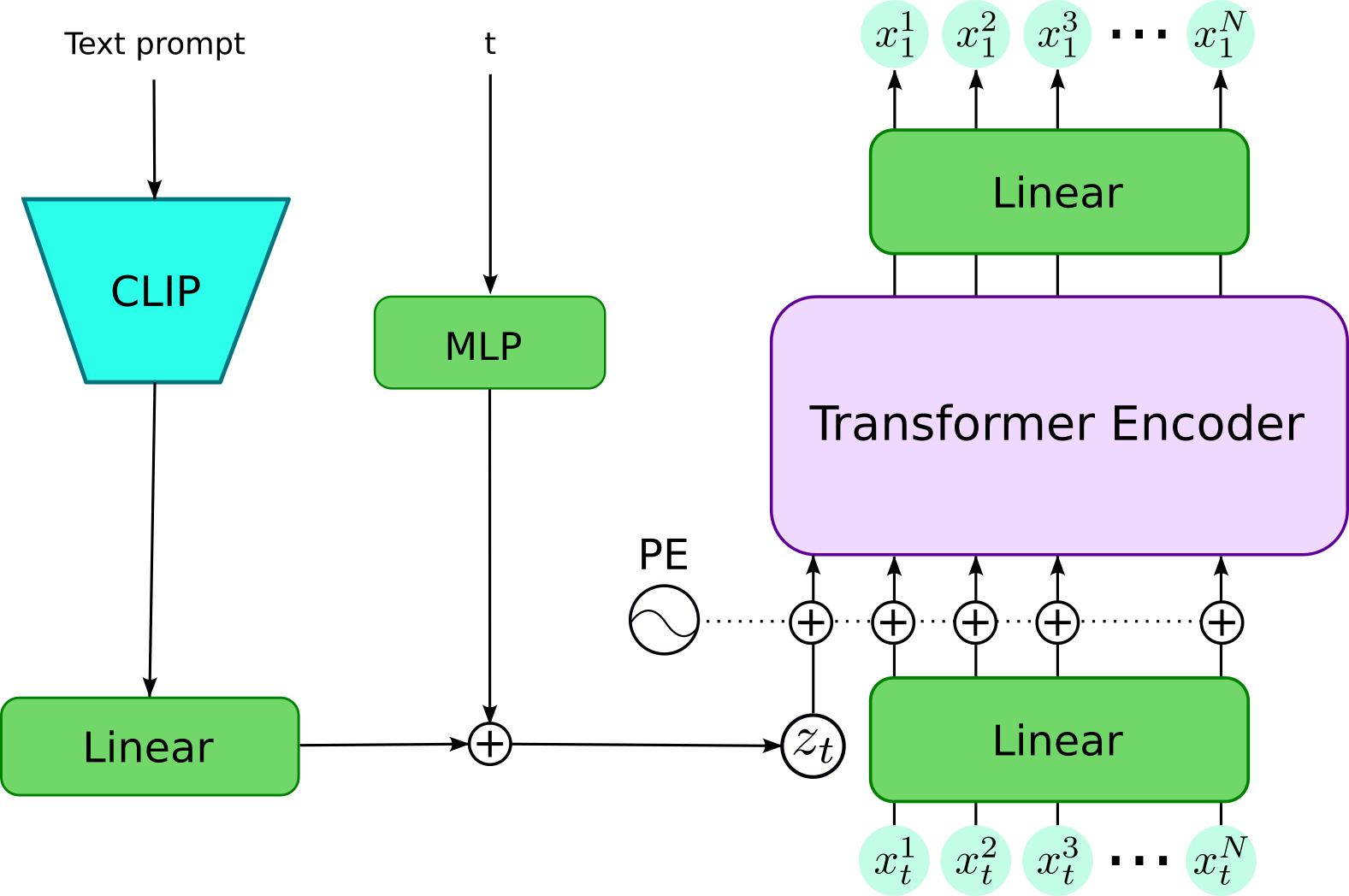} 
    \caption{Text-Driven Motion Generation: Leveraging the architecture proposed by Tevet et al.~\cite{tevet2022human}, our approach generates motion sequences from an input ${x}_t = ({x}^1_t, {x}^2_t, \dots, {x}^N_t)$, where each ${x}^i_t \in \mathbb{R}^{J \times D}$ denotes the pose of the $i$-th frame. Crucially, this input is derived via conditional flow matching. By processing this input through a Transformer Encoder, the model produces a motion sequence ${x}_1 = ({x}^1_1, {x}^2_1, \dots, {x}^N_1)$.}
    \label{fig:model_architecture}
\end{figure}

Crucially, these conditional formulations depend on the samples $x_1$, which in our case correspond to human motion sequences extracted from our datasets. Specifically, $x_1 \in \mathbb{R}^{N \times J \times D}$, where $N$ represents the number of poses in the sequences, $J$ the number of joints per pose, and $D$ the dimensionality of the features for each joint. This entire framework operates under a condition $c$, which, in our context, corresponds to the encoded text (Fig.~\ref{fig:model_architecture}).

While our model implements the architecture of MDM~\cite{tevet2022human}, it is important to highlight that the input $x_t$ is derived from a linear interpolation process, following the CFM framework (Eq.~\eqref{eq:psigauss_expanded}), rather than the stochastic corruption process characteristic of diffusion models such as MDM.

The proposed framework allows us to define a training objective that directly targets motion, thereby mitigating jittering, as outlined in the following.

\subsubsection{Training}\label{sec:training}

\begin{figure}[h]
    \centering
    \includegraphics[width=0.49\textwidth]{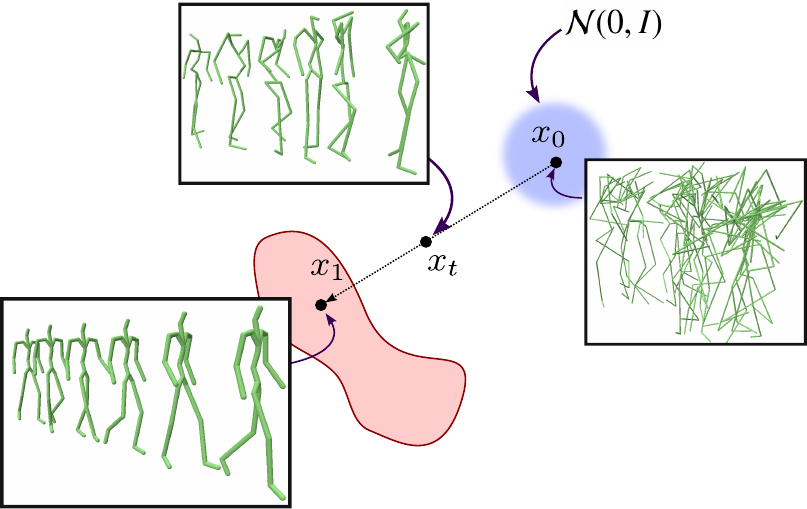}
    \caption{Overview of the training process. In each epoch, the process starts with a sample $x_1$ from the training dataset and a sample $x_0 \sim \mathcal{N}(0, {I})$. An intermediate representation $x_t$ is then determined via linear interpolation between $x_1$ and $x_0$. The region highlighted in red denotes the space of valid human motions. Note that at the beginning of the process, $x_t$ can be found outside this space.}
    \label{fig:model_size_jitter}
\end{figure}

The training process begins by taking a sample $x_1$ from our dataset and simultaneously sampling $x_0$ from the initial Gaussian distribution. We then utilize the flow defined in Eq.~\eqref{eq:psigauss_expanded} to transport $x_0$ along a trajectory conditioned by $x_1$, enabling us to compute the corresponding point $x_t$ at time $t \in [0,1]$. Specifically:
\begin{equation*}
    x_t = \psi_t(x_0) = (1-(1-\sigma_{min})t)x_0 + tx_1,
    \label{eq:flow_equation}
\end{equation*}
where $x_0 \sim \mathcal{N}(0,I)$ (Fig.~\ref{fig:model_size_jitter}).

Building upon the findings of Ramesh et al.~\cite{ramesh2022hierarchical}, which demonstrated that directly predicting the target $x_1$ improved performance over noise prediction—a strategy subsequently adopted by the diffusion-based MDM—we adapt this approach to our Conditional Flow Matching framework. Employing a similar network architecture as MDM, our training objective is thus defined as:

\begin{equation}\label{eq:proxy}
    \mathbb{E}_{x_1\sim q(x_1|c),t\sim \mathcal{U}[0,1]}\lVert \mathcal{G}_p(x_t, t, c; \theta) - x_1  \lVert^{2}_2,
\end{equation}
where $\mathcal{U}[0,1]$ denotes a uniform distribution, $\mathcal{G}_p$ is our trainable model, and $\theta$ represents its learnable parameters. 

Empirically, this formulation reduces motion jitter and improves generalization compared to prior CFM and diffusion-based methods, as validated in Sec.~\ref{sec:results}.

\subsection{Model} 
Our model leverages a Transformer Encoder for motion processing, a technique proven effective for human motion synthesis \cite{petrovich2021action, adewole2024humanmotionsynthesisdiffusion,singleshot}. Adopting the architecture of Tevet et al.~\cite{tevet2022human}, our approach modifies the source of the input $x_t$. Specifically, we employ flow matching to generate $x_t$, while maintaining three key inputs:
\begin{itemize}  
    \item Textual descriptions encoded via a frozen CLIP model.  
    \item Time embeddings through learned positional encoding.  
    \item The motion sequence $x_t$, projected to the latent dimension of the Transformer.  
\end{itemize}  

Text and time encodings are first summed in the latent space, then concatenated with $x_t$ to form the combined input representation (Fig.~\ref{fig:model_architecture}). This fused input is processed by the Transformer to predict the target motion $x_1$ via the objective function (Eq.~\eqref{eq:proxy}), maintaining the temporal resolution of the input while removing noise artifacts. The end-to-end design enables precise control over motion smoothness while preserving fidelity to text conditions. 

\setlength{\tabcolsep}{3pt}
\begin{table*}[ht]
\centering
\caption{Comparison of methods on the HumanML3D dataset. \textcolor{red}{Red} highlights the best values, and \textcolor{blue}{blue} indicates the second-best values. Arrows indicate the desired direction: $\uparrow$ higher is better, $\downarrow$ lower is better, $\rightarrow$ closer to the target value is better.}

\label{table:humanml}
\begin{tabular}{lccccccccc}
\toprule
\multirow{2}{*}{Methods} & 
\multirow{2}{*}{\#params $\downarrow$} & 
\multirow{2}{*}{FID $\downarrow$} & 
\multirow{2}{*}{Diversity $\uparrow$} & 
\multirow{2}{*}{Jitter $\rightarrow$} &
\multirow{2}{*}{FIDJ $\downarrow$} &
\multirow{2}{*}{MM-Dist $\downarrow$} & 
\multirow{2}{*}{MMModality $\uparrow$} &
\multirow{2}{*}{RP Top3 $\uparrow$} \\ 
& & & & & & & & \\
\midrule 
Real motion & 
- &
$0.001^{\pm .000}$ & 
$9.488^{\pm .081}$ & 
$\num{47.262}^{\pm \expon{.069}}$ &  
$0.000$ & 
$2.973^{\pm .005}$ &
- &              
$0.797^{\pm .001}$\\
\midrule 
TM2T \cite{guo2022tm2t} & 
43.0M & 
$1.507^{\pm .019}$ & 
$8.526^{\pm .086}$ & 
$\num{87.208}^{\pm \expon{.151}}$ & 
$16.542$ & 
$3.469^{\pm .011}$ & 
$2.540^{\pm .063}$ & 
$0.726^{\pm .002}$ \\ 

MDM \cite{tevet2022human} & 

\textcolor{red}{17.9M} & 
$0.519^{\pm .050}$ & 
$9.442^{\pm .070}$ & 
\textcolor{red}{$46.30^{\pm .9}$} & 
$5.084$ &
$3.658^{\pm .022}$ & 
\textcolor{blue}{$2.871^{\pm .047}$} & 
$0.706^{\pm .004}$ \\

MFM~\cite{hu2023motion} & 

\textcolor{red}{17.9M} &
$0.446 ^ {\pm .047}$ & 
\textcolor{red}{$ 9.987 ^ {\pm .089}$} & 
$\num{576.679} ^ {\pm \expon{3.153}}$ & 
$32.906$ & 
$5.186 ^ {\pm .023}$ & 
$2.419 ^ {\pm .096}$ & 
$0.666 ^ {\pm .005}$ \\

MLD \cite{mld} & 

26.4M & 
$0.427 ^ {\pm .012}$ & 
$9.775 ^ {\pm .073}$ & 
$\num{22.772} ^ {\pm \expon{.113}}$ & 
$3.391$ &
$3.269 ^ {\pm .015}$ & 
$2.573 ^ {\pm .089}$ & 
$0.758 ^ {\pm .003}$ \\

MotionLCM \cite{dai2024motionlcm} & 

39.7M & 
$0.416 ^ {\pm .010}$ & 
$9.740 ^ {\pm .069}$ & 
$\num{37.578} ^ {\pm \expon{.224}}$ & 
$3.742$ & 
\textcolor{red}{$3.008 ^ {\pm .005}$} & 
$2.123 ^ {\pm .076}$ &
\textcolor{red}{$0.802 ^ {\pm .002}$} \\ 
MotionGPT \cite{jiang2023motiongpt}& 

332.8M & 
\textcolor{blue}{$0.185 ^ {\pm .007}$} & 
$9.291 ^ {\pm .067}$ &
$\num{94.770} ^ {\pm \expon{.250}}$ &
$4.148$ & 
$4.013 ^ {\pm .022}$ & 
\textcolor{red}{$3.481 ^ {\pm .119}$} & 
$0.659 ^ {\pm .003}$ \\

T2M-GPT \cite{zhang2023generating} & 

 228.4M & 
\textcolor{red}{$0.124^{\pm .008}$} & 
$9.567^{\pm .101}$ &
$\num{76.046}^{\pm \expon{1.123}}$ &
$2.615$ & 
\textcolor{blue}{$3.131^{\pm .019}$} &
$1.719^{\pm .120}$ & 
\textcolor{blue}{$0.776^{\pm .005}$} \\ 
\midrule
FlowMotion-Big& 
\textcolor{blue}{26.3M} & 
$0.268 ^ {\pm .028}$ & 
\textcolor{blue}{$9.796 ^ {\pm .069}$} & 
\textcolor{blue}{$40.81 ^ {\pm .5}$} &
\textcolor{red}{$2.400$} &
$5.319 ^ {\pm .032}$ &
$2.327 ^ {\pm .077}$ & 
$0.648 ^ {\pm .004}$ \\ 

FlowMotion& 
\textcolor{red}{17.9M} & 

$0.278 ^ {\pm .030}$&
$9.788 ^ {\pm .103}$ & 
$\num{39.522} ^ {\pm \expon{.770}}$ & 
\textcolor{blue}{$2.453$} & 
$5.239 ^ {\pm .027}$ &
$2.349 ^ {\pm .055}$ &
$0.663 ^ {\pm .005}$ & 

\end{tabular}
\end{table*}
\subsection{Sampling}
To generate the final point $x_1$ at time $t=1$, the sampling process begins with $x_0 \sim \mathcal{N}(0,I)$ at time $t=0$. We then propagate $x_0$ through the flow $\psi$, as defined in Sec.~\ref{sec:training}. This involves solving Eq.~\eqref{eq:ode} with initial condition $\psi_0(x_0) = x_0$ to obtain $\psi_1(x_0)$. We approximate the solution via Euler integration~\cite{hu2023motion}, yielding the iterative update:  

\begin{equation}
\label{eq:eulerfield}
x_{t_{i+1}} = x_{t_{i}} + h \left( u_{t_i}(x_{t_i}|x_1) \right),
\end{equation}
where $i \in \{0, 1, \dots, M - 1\}$, $M$ denotes the number of iterations, $h=1/M$ represents the step size, and the time instance is given by  $t_i=i/M$. Observe that $t_i \in [0,1)$. The conditional vector field $u_{t_i}$, defined in Eq.~\eqref{eq:campotheorema},  depends on  $x_1$ and $x_{t_i}$.

To enhance the influence of the textual condition $c$, we introduce a guided model $\mathcal{G}$ based on the classifier-free guidance technique. Leveraging our trained model $\mathcal{G}_p$, which predicts $x_1$, the guided model $\mathcal{G}$ is defined as:
\begin{equation*}
\begin{split}
\mathcal{G}(x_t, t, c; \theta) &= \mathcal{G}_p(x_t, t, \emptyset; \theta) \\
&+ s \left( \mathcal{G}_p(x_t, t, c; \theta) - \mathcal{G}_p(x_t, t, \emptyset; \theta) \right)
\end{split}
\label{eq:guidance}
\end{equation*}
where $s$ is the guidance scale. This formulation interpolates between conditional (with text input $c$) and unconditional (where $c = \emptyset$) generation, allowing us to control the influence of the textual condition. 

Thus, substituting $\mathcal{G}$ into Eq.~\eqref{eq:campotheorema} and replacing into Eq.~\eqref{eq:eulerfield}, we obtain the iterative sampling process:
\begin{equation*}
    x_{t_{i+1}} = x_{t_i} + h \Bigg[ \frac{\mathcal{G}(x_{t_i}, t_i, c; \theta) - (1-\sigma_{\text{min}})x_{t_i}}{1 - (1-\sigma_{\text{min}})t_i}\Bigg].
    \label{eq:sampling_update}
\end{equation*}

As previously mentioned, the objective of flow matching, and in our specific case CFM, is to determine the conditional vector field $u_{t}$. This is achieved from the model designed to estimate the target directly. By obtaining $\mathcal{G}$ and having the previous iterative value $x_{t_i}$, we can derive the desired vector field using Eq.~\eqref{eq:campotheorema} without the need to generate the vector field directly. 

\subsection{Implementation Details}
\label{sec:implementation}
The Transformer Encoder architecture utilizes a hidden dimension of 512. The text encoder comprises 8 layers, 4 attention heads, and a feedforward hidden layer dimension of 1024. We employ the pretrained CLIP-ViT-B/32 model for text processing, keeping its parameters frozen throughout training. Our model was trained with a batch size of 128.

In our implementation, we set $\sigma_{min}$ to zero during both sampling and training. While theoretically, the standard deviation $\sigma_{min}$ of a Gaussian distribution must be strictly positive, empirically, setting $\sigma_{min} = 0$ yielded improved performance in our experiments.

We use Euler integration with 100 steps and a classifier-free guidance scale of 2.5, a standard value in similar works~\cite{tevet2022human}. For optimization, we adopted the AdamW~\cite{loshchilov2017decoupled} optimizer, configured with parameters $\beta_1$ = 0.9 and $\beta_2$ = 0.999, and a learning rate set to $1 \times 10^{-4}$.

\section{Experiments}\label{secexperiments}
Experiments were conducted on a Linux system using NVIDIA A30 and RTX 3060 GPU cards. The A30 GPU was used for training, which took approximately 24 hours. The RTX 3060 GPU was used for evaluating metrics and comparing against state-of-the-art methods. Sample generation for our model, performed on the RTX 3060, required approximately 2 seconds per instance. In contrast, the MDM~\cite{tevet2022human}, which is based on diffusion, required approximately 30 seconds for sample generation on the same RTX 3060 GPU. This speed constitutes a significant improvement and also surpasses other diffusion-based models in sampling speed~\cite{MotionDiffuse}.

Our model generates human motion sequences conditioned on textual descriptions. To quantitatively assess the quality and diversity of the generated motions, we adopt a range of metrics proposed by Guo et al.~\cite{Guo_2022} for evaluating text-conditioned human motion generation. These metrics allow us to compare our approach to state-of-the-art methods across various dimensions, including the fidelity of the motion to the text, the diversity of the generated motions, and the smoothness of the motion.

\subsection{Datasets}
We evaluate FlowMotion on two benchmark datasets: HumanML3D~\cite{Guo_2022} and KIT~\cite{kitml}.

\begin{itemize}
    \item \textbf{HumanML3D}: This dataset, introduced by Guo et al.~\cite{Guo_2022}, combines the HumanAct12~\cite{guo2020action2motion} and AMASS~\cite{mahmood2019amass} datasets. It consists of 14,616 motion sequences at 20 frames per second. The original motion sequences can exceed 10 seconds in duration, but following prior work~\cite{Guo_2022}, they are randomly truncated to a maximum of 10 seconds (196 frames) for consistency, with a minimum sequence length of 40 frames. The dataset is paired with 44,970 textual descriptions.  The results of our experiments on this dataset are detailed in Table~\ref{table:humanml}.    
    \item \textbf{KIT Motion-Language Dataset}~\cite{kitml}: This dataset provides a valuable resource for evaluating text-conditioned motion generation, containing 3,911 motion sequences and 6,353 textual descriptions that capture a diverse range of human actions and movements. Motion sequences have a maximum length of 196 frames and a minimum of 24 frames. The results for this dataset are presented in Table~\ref{table:kitml}.
\end{itemize}
\begin{table*}[ht]
\centering
\caption{Comparison of methods on the KIT dataset. \textcolor{red}{Red} highlights the best values, and \textcolor{blue}{blue} indicates the second-best values. Arrows indicate the desired direction: $\uparrow$ higher is better, $\downarrow$ lower is better, $\rightarrow$ closer to the target value is better.}
\label{table:kitml}
\begin{tabular}{lccccccccc}
\toprule
\multirow{2}{*}{Methods} & 
\multirow{2}{*}{\#params $\downarrow$} &
\multirow{2}{*}{FID $\downarrow$} & 
\multirow{2}{*}{Diversity $\uparrow$} & 
\multirow{2}{*}{Jitter $\rightarrow$} &
\multirow{2}{*}{FID-J $\downarrow$} &
\multirow{2}{*}{MM-Dist $\downarrow$} & 
\multirow{2}{*}{MMModality $\uparrow$} &
\multirow{2}{*}{RP Top3 $\uparrow$} \\ 
& & & & & & & \\
\midrule 
Real motion & 
- & 
$0.026^{\pm .003}$ & 
$11.016^{\pm .095}$ & 
$\num{49.922}^{\pm \expon{.165}}$ & 
$0.00$ & 
$2.772^{\pm .013}$ & 
- &                
$0.784^{\pm .003}$ \\
\midrule 
T2M-GPT \cite{zhang2023generating} &
\textcolor{blue}{228.4M} & 
$0.469^{\pm .010}$ & 
\textcolor{red}{$11.006^{\pm .111}$} &
$\num{98.425} ^ {\pm \expon{.291}}$ & 
\textcolor{blue}{$8.729$} &
\textcolor{red}{$3.002^{\pm .013}$} & 
\textcolor{red}{$1.903 ^{\pm .070}$} &  
\textcolor{red}{$0.737^{\pm .004}$} \\
MDM \cite{tevet2022human} & 
\textcolor{red}{17.9M} & 
$0.505^{\pm .027}$ &
$10.705^{\pm .098}$ & 
\textcolor{blue}{$73.21 ^ {\pm .5}$} &
$10.196$ & 
\textcolor{blue}{$3.077^{\pm .017}$} & 
\textcolor{blue}{$1.782 ^ {\pm .152}$} & 
\textcolor{blue}{$0.731^{\pm .004}$} \\ 
MFM~\cite{hu2023motion} & 
\textcolor{red}{17.9M} & 
\textcolor{red}{$0.327 ^ {\pm .016}$} & 
$10.707 ^ {\pm .070}$ & 
$\num{1099.988} ^ {\pm \expon{1.183}}$ &
$53.801$ & 
$9.155 ^ {\pm .028}$ &
$1.639 ^ {\pm .137}$ & 
$0.405 ^ {\pm .004}$ \\
\midrule
FlowMotion & 

\textcolor{red}{17.9M} & 
\textcolor{blue}{$0.396 ^ {\pm .042}$} &
\textcolor{blue}{$10.989 ^ {\pm .082}$} & 
\textcolor{red}{$52.40 ^ {\pm .3}$} &
\textcolor{red}{$8.362$} &
$9.206 ^ {\pm .022}$ & 
$1.756 ^ {\pm .083}$ & 
$0.404 ^ {\pm .005}$ \\ 
\end{tabular}
\end{table*}
\subsection{Motion Representation}
\label{subsec:motionrepresentation}
Consistent with Guo et al.~\cite{Guo_2022}, we represent each pose using the features $x_f = (\dot{r}_a, \dot{r}_x, \dot{r}_z, r_y, j_p, j_v, j_r, c_f)$, where $\dot{r}_a \in \mathbb{R}$ denotes the global root angular velocity, $\dot{r}_x, \dot{r}_z \in \mathbb{R}$ represent the global root velocity in the X-Z plane, $r_y$ is the root height, $j_p \in \mathbb{R}^{3j}$, $j_v \in \mathbb{R}^{3j}$, $j_r \in \mathbb{R}^{6j}$ correspond to the local joint positions, velocities, and rotations respectively, with $j$ being the number of joints, and $c_f \in \mathbb{R}^4$ represents the foot contact features, derived from the heel and toe joint velocities. Thus, each frame within our motion sequences is characterized by this set of features, $x_f$, meaning that each $x_t$ input to our model comprises frames with this format.

These motion features encapsulate local joint positions, velocities, and rotations within the root space, in addition to global translation and rotation. The dimensionality of these features is directly determined by the number of joints considered. Specifically, for the HumanML3D dataset, which uses $j=22$ joints, the resulting feature dimension is 263. The KIT-ML dataset, with $j=21$, has a feature dimension of 251. This difference in dimensionality reflects the varying complexity captured by the motion representation of each dataset.

\subsection{Metrics}
\label{sec:metrics}
We employ evaluation metrics suggested by Guo et al.~\cite{Guo_2022}, a standard practice in the human motion generation literature~\cite{zhang2023generating,MotionDiffuse,zhang2023remodiffuse,hu2023motion,tevet2022human}. These metrics include Fréchet Inception Distance (FID), R-Precision (RP Top-k), Multimodal Distance (MM-Dist), Diversity, and Multimodality (MModality). Following established evaluation protocols, each metric, except for MModality, is computed over 20 trials to account for the inherent stochasticity in the sampling and evaluation process. MModality is calculated after generating 30 samples for a same set of randomly chosen batches. We report the average and standard deviation across these trials, as shown in Tables~\ref{table:humanml}-~\ref{table:kitml}. In addition to this established suite, we introduce Jitter, which quantifies motion smoothness, and Mahalanobis FID-J, which measures the balance between fidelity and motion naturalness. As with the other metrics, Jitter is computed over 20 trials to account for stochasticity.

\paragraph{R-Precision (RP Top-k)} Measures the correspondence between the generated motion and the input text. For RP Top-k, batches of 32 (text embedding, motion embedding) pairs from the test set are processed. The distance between each text embedding and all motion embeddings in the batch is calculated. We count how often the correct motion embedding appears among the 1, 2, or 3 nearest neighbors (Top-1, Top-2, Top-3). This is repeated for all batches, and the final percentage is calculated by dividing by the total number of samples.

\paragraph{Multimodal Distance (MM-Dist)}  Calculates the average distance between a set of randomly sampled text embeddings and their corresponding generated motion embeddings.

\paragraph{Diversity}  Assesses the variability of the generated motions. Motions are encoded into a shared latent space, and the average pairwise distance between randomly sampled motions is calculated. A greater average distance signifies higher diversity.

\paragraph{Multimodality (MModality)}  Determines the average Euclidean distance between generated motions encoded in the same latent space when conditioned on the identical text. MModality is calculated after generating 30 samples for the same set of randomly chosen batches.

\paragraph{Params}  reports the total number of parameters in each model's architecture. It reflects the overall size and complexity of the network used during training.

\paragraph{Jitter} Evaluates motion smoothness by measuring the jerk (rate of change of the acceleration) of each joint. Following Du et al.~\cite{du2023avatars}, we compute it as the average magnitude of the acceleration of body joints. Lower jitter values indicate smoother motion. To ensure comparability across datasets, we introduce a scaling factor, denoted as $\alpha$, derived from the ratio of motion ranges of the HumanML3D and KIT datasets. The motion range for each dataset is calculated as the difference between the mean of the maximum joint positions and the mean of the minimum joint positions across all motion sequences. Formally, the scaling factor $\alpha$ is computed as:
\begin{equation*}
\alpha = \frac{\text{range}_{\text{humanml3d}}}{\text{range}_{\text{kit}}} = \frac{\bar{x}^{\text{max}}_{\text{humanml3d}} - \bar{x}^{\text{min}}_{\text{humanml3d}}}{\bar{x}^{\text{max}}_{\text{kit}} - \bar{x}^{\text{min}}_{\text{kit}}},
\end{equation*}
where $\bar{x}^{\text{max}}_{d}$ and $\bar{x}^{\text{min}}_{d}$ represent the mean of the maximum and minimum joint positions for dataset $d$, respectively. Empirically, $\alpha$ was determined to be 0.00073. This scaling factor is applied to modulate the magnitude of jitter values obtained from the KIT dataset using the formula:
\begin{equation*}
\text{Jitter}_{\text{kit}}^{\text{scaled}} = \alpha \cdot \text{Jitter}_{\text{kit}}.
\end{equation*} 

This normalization mitigates the tendency for jitter values to become excessively large on the KIT dataset, ensuring a consistent perturbation magnitude across datasets and enhancing robustness and comparability.

To facilitate the comparison of textual and motion data, both
are encoded into a shared latent space. This is enabled by the
pretrained motion and text encoders provided by Guo et al.~\cite{Guo_2022},
allowing for quantitative evaluation based on Euclidean distances in this embedding space.

\paragraph{Mahalanobis FID-J} We introduce the Mahalanobis FID-J metric to quantify the balance between FID and motion jitter. It measures the Mahalanobis distance, defined as:

\begin{equation}
    D_M(x) = \sqrt{(x - \mu)^T S^{-1} (x - \mu)},
    \label{eq:mahalanobis}
\end{equation}

of a method's FID and jitter vector ($x$) from the ground truth FID and jitter vector ($\mu$) in the FID-Jitter space. In this context, $S$ is the covariance matrix. The ground truth values represent the ideal performance point, and minimizing the Mahalanobis distance implies achieving an optimal balance between these two metrics. Given that FID and Jitter have different scales (Fig.~\ref{fig:jittercompare}), the Mahalanobis distance, a statistical measure that accounts for the covariance between the metrics, is more appropriate than the Euclidean distance. This allows us to effectively evaluate how well a method jointly minimizes both FID and jitter.
    
To ensure a fair comparison of methods, outlier detection was applied to the FID and Jitter values solely for the purpose of computing the covariance matrix ($S$ in Equation \ref{eq:mahalanobis}), particularly due to the presence of high Jitter values for the MFM method (576.68 and 1099.99) in Table~\ref{table:humanml} (HumanML3D) and Table~\ref{table:kitml} (KIT), respectively.

\begin{itemize}
    \item For HumanML3D (Table~\ref{table:humanml}), outlier detection was performed using the Mahalanobis distance, identifying the TM2T method (FID=1.507, Jitter=87.21) and the MFM method (FID=0.446, Jitter=576.68) as the outlier.
    \item For the KIT dataset (Table~\ref{table:kitml}), given the limited number of evaluated models (four), we employed a robust z-score based on the Median Absolute Deviation (MAD), applied specifically to the jitter values to identify potential outliers. This identified the MFM method with a Jitter value of 1099.99 as an outlier. Values with $|z| > 3$ were considered outliers.
\end{itemize}

For both HumanML3D (Table~\ref{table:humanml}) and KIT (Table~\ref{table:kitml}), after computing the covariance matrix (after outlier removal if applicable), this matrix was then used to calculate the Mahalanobis distance between all methods and the ground truth reference, using Equation \ref{eq:mahalanobis}, allowing for a more robust and reliable comparison across methods.

\subsection{Model Parameter Analysis}
\label{sec:model_parameters}
Our model is based on a Transformer Encoder, as introduced by Vaswani et al.~\cite{allattention}. These networks allow for flexibility in adjusting parameters like the feedforward dimension and the number of attention heads. Drawing inspiration from the original Transformer work, a scaled-up version is also explored. We introduce two variants:
\begin{itemize}
    \item \textbf{FlowMotion (Standard):} 8 layers, 4 attention heads, 1024 feedforward dimension
    \item \textbf{FlowMotion-Big:} 8 layers, 16 attention heads, 2048 feedforward dimension
\end{itemize}

The scaled-up FlowMotion-Big variant improves the FID-Jitter balance marginally  (Table~\ref{table:humanml}); however, this comes at the cost of increased the number of parameter of the model. Both variants retain the original Transformer 512-dimensional hidden representation in the encoder, consistent with prior work~\cite{allattention}.

\begin{figure}[!ht]
    \centering
    \includegraphics[width=0.415\textwidth]{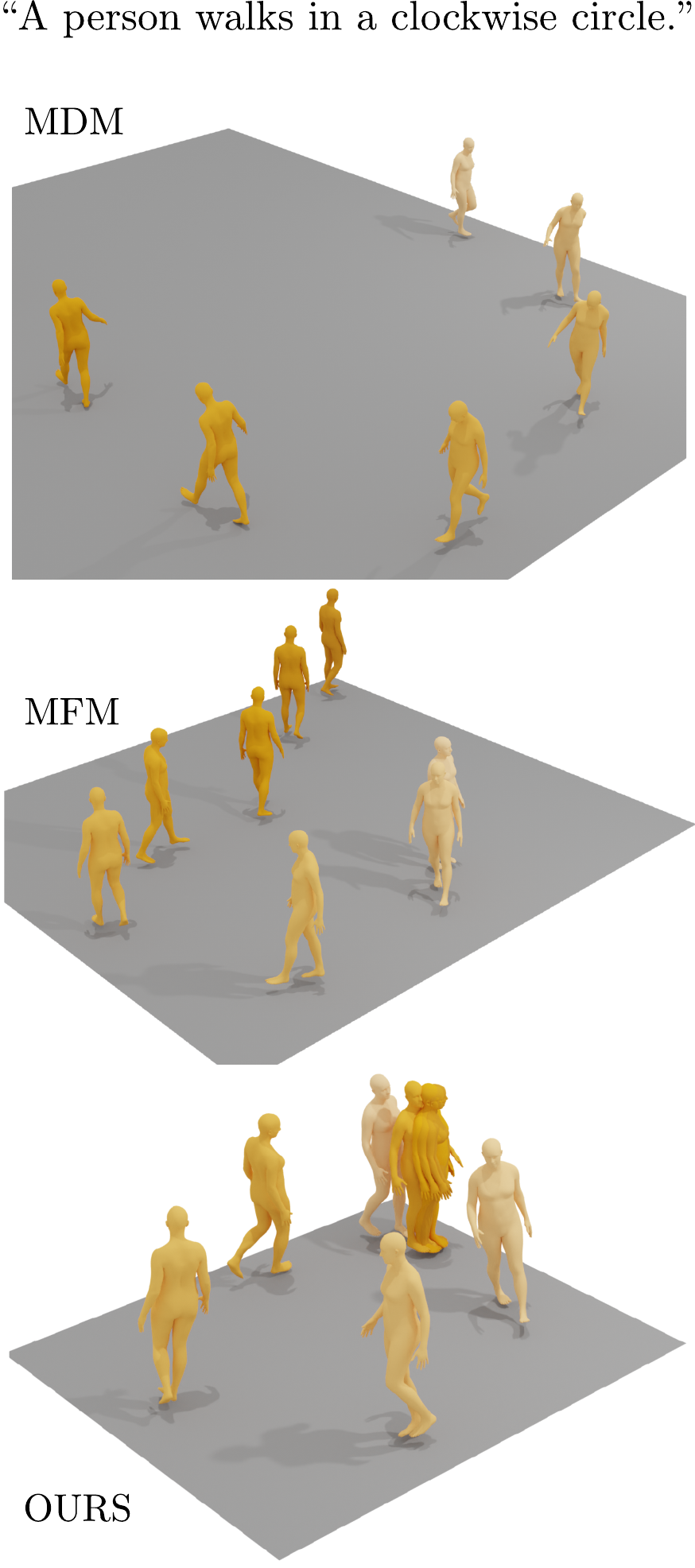}
    \caption{
Comparison of trajectory generation performance between MDM~\cite{tevet2022human}, MFM~\cite{hu2023motion}, and the FlowMotion Model for circular motion instructions. MDM and MFM exhibit significant errors in trajectory closure. The FlowMotion Model, conversely, generates a trajectory with accurate start and end point coincidence, indicative of superior motion instruction understanding. The FlowMotion Model's improved FID score confirms its enhanced generation fidelity to the intended circular motion.}
    \label{fig:results}
\end{figure}

\section{Results}
\label{sec:results}
Our model demonstrates a superior trade-off between fidelity and motion smoothness, achieving notably low Jitter values while maintaining high-quality motion generation. On the KIT dataset (Table~\ref{table:kitml}), it achieves the best Jitter reduction with a score of 52.4, close to the ground truth value of 49.9. On HumanML3D (Table~\ref{table:humanml}), it achieves a Jitter score of 39.52, outperforming MDM (46.3) and approaching the ground truth (47.26). While MDM shows a slightly better Jitter on HumanML3D, it suffers from significantly higher FID and a substantially slower sampling speed (30 seconds vs. our 2 seconds). 

To better capture the trade-off between fidelity and naturalness, we introduce the FIDJ metric, demonstrating that we achieve an optimal balance on both datasets. This metric measures the Mahalanobis distance from each method’s (FID, Jitter) pair to the ground truth point: (0.001, 47.26) for HumanML3D and (0.026, 49.92) for KIT. In this formulation, the ground truth naturally has a FIDJ of 0, as it defines the reference point. Our model achieves FIDJ scores of 2.4 on HumanML3D and 8.3 on KIT, indicating that it remains closest to the optimal point in the FID–Jitter space. These results highlight the model’s effectiveness for applications requiring both realistic and high-fidelity human motion generation.
\begin{figure}[!htbp]
    \centering
    \includegraphics[width=0.50\textwidth]{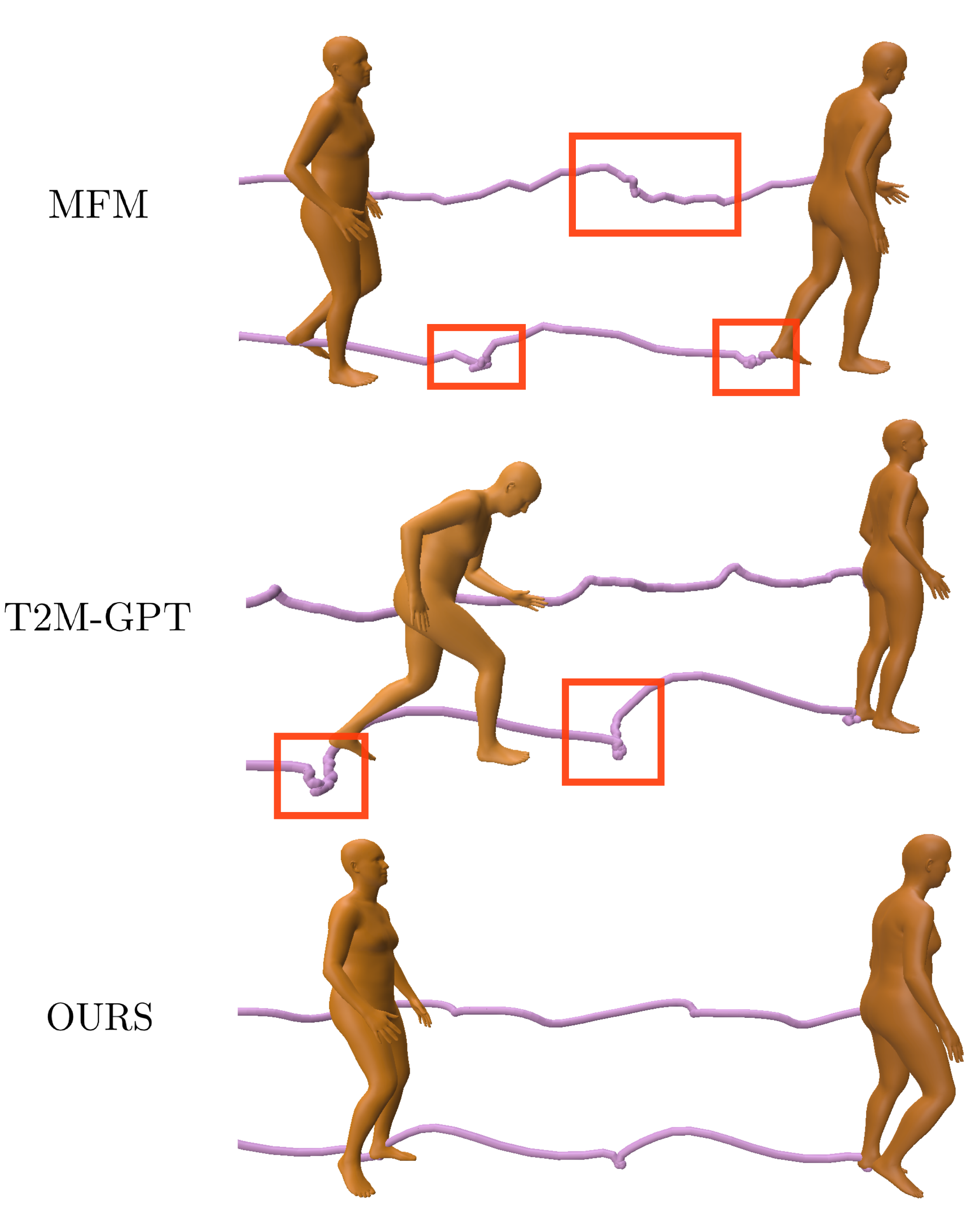}
    \caption{Qualitative comparison of motion sequences generated from the text prompt ``a person walks forward in a straight line", comparing MFM~\cite{hu2023motion}, T2M-GPT~\cite{zhang2023generating}, and FlowMotion. The figure displays multiple poses of the character, spaced in time, with wrist and heel trajectories shown to visualize the motion path.}
    \label{fig:smoothcompare}
\end{figure}

On the HumanML3D dataset, our model—built upon the same underlying architecture as MDM and MFM—achieves a lower FID score than both (Fig.~\ref{fig:results}). In contrast, models using Vector Quantized Variational Autoencoders (VQ-VAE), such as T2M-GPT and MotionGPT, exhibit higher fidelity (FID) but tend to produce less smooth motion sequences, with Jitter values approximately double that of the ground truth (Tables~\ref{table:humanml}-\ref{table:kitml}). The superior smoothness is also visually apparent when comparing the generated motions, as shown in Fig.\ref{fig:smoothcompare}, where trajectories of body parts over time are observed. As shown in Fig.\ref{fig:smoothcompare}, MFM and T2M-GPT produce motions with noticeable jerkiness. Our method exhibits smoother transitions throughout the motion, making it appear more natural, underscoring its effectiveness in generating temporally smooth motion.

\begin{figure}[!htbp] 
    \centering
    \subfloat[]
    {\label{fig:temblores_mfm}
        \includegraphics[width=0.45\textwidth]{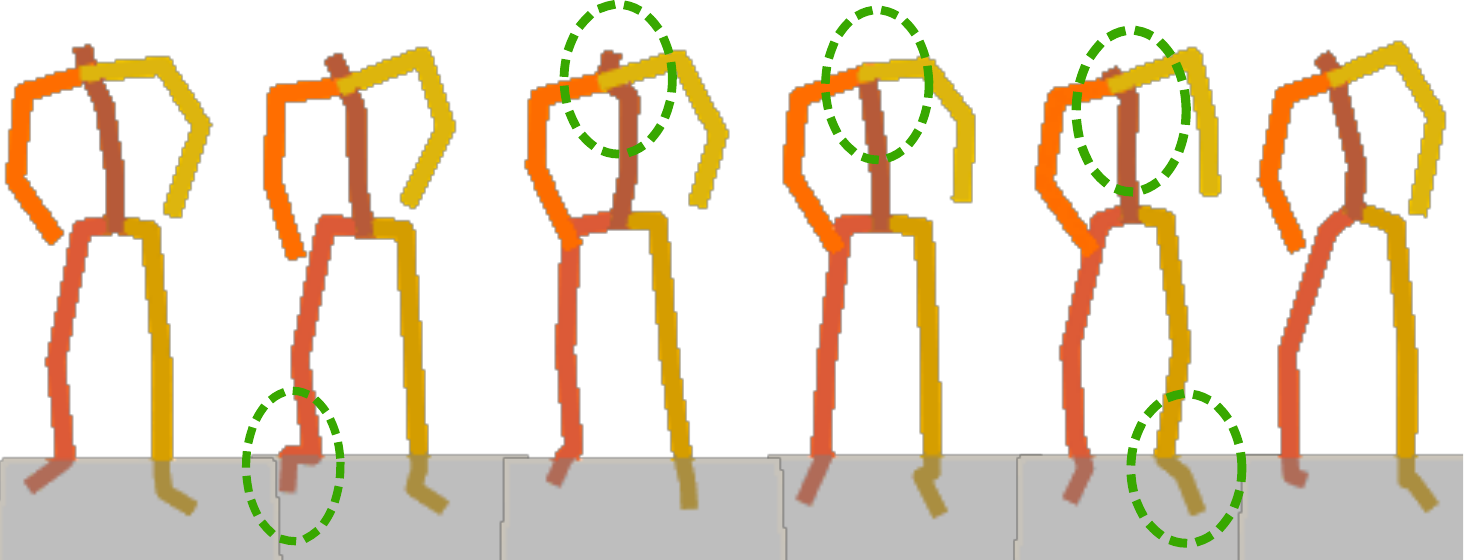}
    }
    \\[1ex]
    \subfloat[]
    {\label{fig:temblores_ours}
        \includegraphics[width=0.46\textwidth]{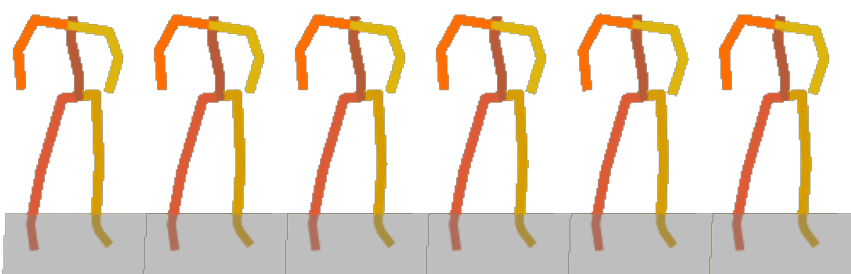}
    }
    \caption{Motion generation comparison for KIT-ML trained models with prompt ``a person walks forwards and stops'' at pre-step instant: (a) MFM: tremors and motion artifacts including body part distortions, visible within green demarcations (b) FlowMotion: stable and coherent pose}
    \label{fig:comparetemblores}
\end{figure}

On the KIT dataset, our method achieves a low FID-J (8.3) that outperforms all other methods in Table \ref{table:kitml}, highlighting its ability to generate smooth, high-quality motion sequences while maintaining excellent fidelity. Further evidence of robustness of our model, when trained on KIT-ML~\cite{kitml}, is presented in Figure~\ref{fig:comparetemblores}, directly comparing motion generation against MFM~\cite{hu2023motion}. A key advantage, enhanced motion stability, is visually highlighted: Figure~\ref{fig:temblores_mfm} shows MFM generation with noticeable tremors and motion artifacts, including body part instability, even during intended stillness before a step. In contrast, captured at the same instant (Figure~\ref{fig:temblores_ours}), our FlowMotion model showcases stability, maintaining a coherent pose without involuntary movements or tremors. This visual comparison underscores improved stability and reduced artifact generation achieved by FlowMotion.

\subsection{Ablation Study}
Regarding the ablation study, it is important to note that the original work of MFM achieved smoother results, albeit with a larger batch size (256). Due to hardware limitations, we were constrained to use a batch size of 128. Nevertheless, MFM represents the process of using flow matching training to determine the vector field; hence, it serves as our ablation study. In our case, we perform the training by directly determining the value without noise $x_1$. 
The results demonstrate the advantages of our direct training approach over the conventional flow-matching technique, even under similar training conditions.

\begin{figure}[!htbp]
    \centering
    \includegraphics[width=0.45\textwidth]{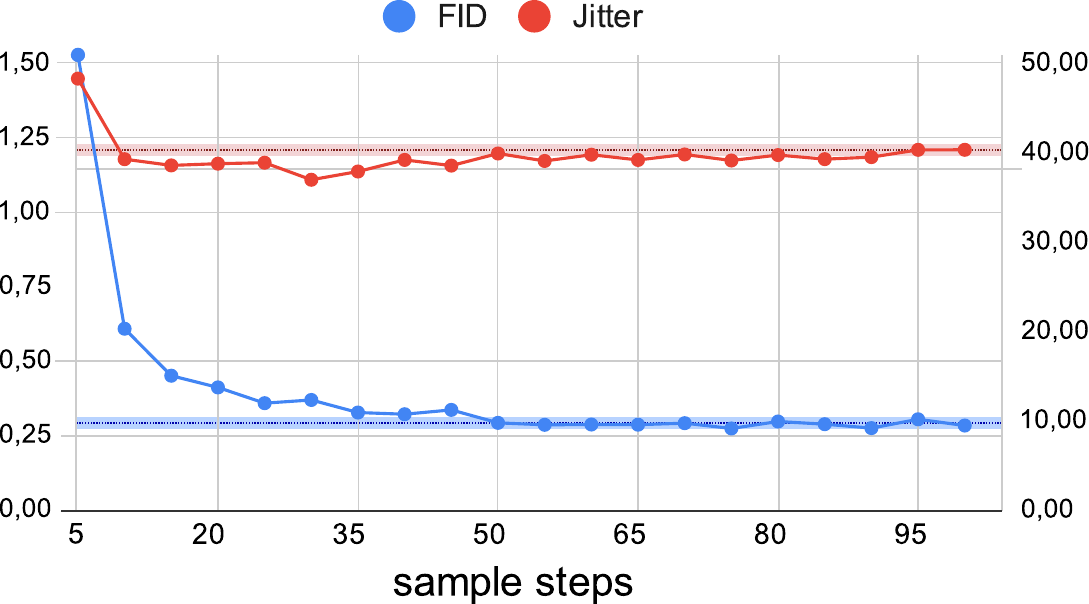}
    \caption{Comparison of FID and Jitter for our model across 100 sampling steps, taken every 5 steps. The graph displays results from the 5th step onwards for better visualization. The horizontal blue band represents the FID value of 0.278 $\pm$ 0.03, and the horizontal red band represents the Jitter value of 39.5 $\pm$ 0.77, both obtained by our model as reported in Table~\ref{table:humanml}.}
    \label{fig:fid_jitter}
\end{figure}
Our model also performs strongly on Diversity, achieving scores among the top two on both the HumanML3D (Table~\ref{table:humanml}) and KIT (Table~\ref{table:kitml}) datasets, suggesting that it explores the motion space in a rich and varied manner. Conversely, it shows lower performance on the RP Top3, MM-Dist, and MM-Modality metrics, indicating a less strict adherence to the text condition. This can be attributed to the model's focus on preserving realism and diversity rather than replicating canonical or literal motions. We consider this trade-off consistent with the model's objectives, which aim to offer more fluid, natural, and diverse generations, even if they are less deterministic.

We also analyze the trade-off between motion quality and sampling steps, as illustrated in Fig.~\ref{fig:fid_jitter}, which shows that both FID and Jitter values exhibit a trend toward stabilization after approximately 50 sampling steps, approaching the error region of our previously reported results (horizontal lines in Fig.~\ref{fig:fid_jitter}). A more complete set of metrics, including diversity and R-precision, is presented in Table~\ref{table:stepshumanml} for sampling steps ranging from 1 to 100. Notably, all metrics remain relatively stable from around 50 steps onward, consistent with our results in Table~\ref{table:humanml}. This analysis, conducted on the HumanML3D test set, indicates that a sampling regime of around 50 steps is sufficient for generating high-quality motions with good fidelity, striking a balance between computational cost and motion fidelity.

\begin{figure}[!htbp]
    \centering
    \includegraphics[width=0.45\textwidth]{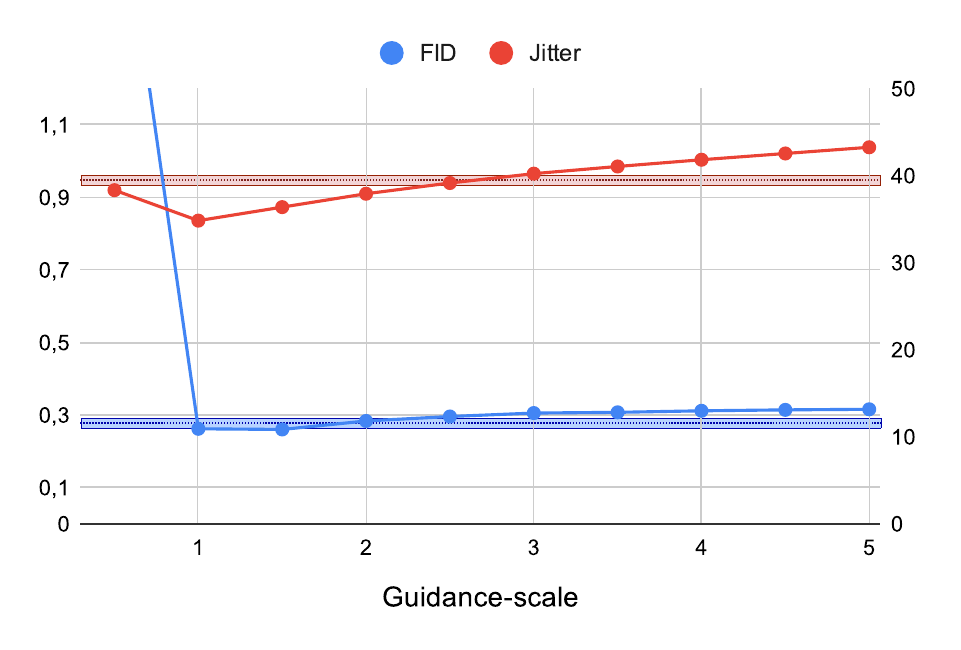}
    \caption{FID and Jitter metrics across varying guidance scales for our model, ranging from 0.5 to 5 in increments of 0.5.}
    \label{fig:guidance}
\end{figure}

Further analysis of our model's performance under various guidance strengths reveals an optimal balance at a guidance scale of 2.5 (Fig.~\ref{fig:guidance}). At this setting, the model maintains an FID score below 0.3 while achieving a Jitter score near 40, approaching the ground truth value of 47. This configuration, which we employed for our model's generation, demonstrates the effectiveness of carefully tuning the guidance scale to achieve both high fidelity and smooth motion quality.

\begin{table*}[!htbp]
\centering
\caption{Comparison of metrics for different numbers of sampling steps (NS) on the HumanML3D dataset. \textcolor{red}{Red} highlights the best values, and \textcolor{blue}{blue} indicates the second-best values. Arrows indicate the desired direction: $\uparrow$ higher is better, $\downarrow$ lower is better, $\rightarrow$ closer to the target value is better. }
\label{table:stepshumanml}
\begin{tabular}{lccccccc}
\toprule
\multirow{2}{*}{NS} & 
\multicolumn{3}{c}{R-Precision $\uparrow$} & 
\multirow{2}{*}{FID $\downarrow$} & 
\multirow{2}{*}{MM-Dist $\downarrow$} & 
\multirow{2}{*}{Diversity $\uparrow$} & 
\multirow{2}{*}{Jitter $\rightarrow$} \\
\cmidrule(lr){2-4}
& Top-1 & Top-2 & Top-3 & & & & \\

\midrule 
1 & $0.294^{\pm .005}$ & $0.463^{\pm .005}$ & $0.581^{\pm .004}$ & $2.515^{\pm .065}$ & $5.601^{\pm .023}$ & $8.992^{\pm .075}$ & $\num{188.013}^{\pm \expon{1.751}}$ \\
5 & $0.318^{\pm .005}$ & $0.492^{\pm .005}$ & $0.607^{\pm .005}$ & $1.528^{\pm .071}$ & $5.622^{\pm .026}$ & $9.485^{\pm .080}$ & \textcolor{red}{$47.40^{\pm 0.3}$} \\
10 & $0.353^{\pm .006}$ & $0.535^{\pm .009}$ & $0.645^{\pm .007}$ & $0.610^{\pm .033}$ & $5.360^{\pm .032}$ & $9.757^{\pm .094}$ & $\num{38.567}^{\pm \expon{0.300}}$ \\

20 &\textcolor{blue}{ $0.375^{\pm .006}$} & $0.557^{\pm .007}$ & $0.663^{\pm .006}$ & $0.413^{\pm .040}$ & $5.239^{\pm .033}$ & \textcolor{red}{$9.926^{\pm .100}$} & $\num{38.074}^{\pm \expon{0.506}}$ \\

30 & $0.373^{\pm .005}$ & \textcolor{blue}{$0.558^{\pm .006}$} & \textcolor{red}{$0.667^{\pm .005}$} & $0.371^{\pm .037}$ & $5.237^{\pm .029}$ & $9.807^{\pm .086}$ & $\num{36.310}^{\pm \expon{0.611}}$ \\

40 & $0.373^{\pm .006}$ & $0.553^{\pm .007}$ & $0.654^{\pm .006}$ & $0.324^{\pm .041}$ & $5.253^{\pm .033}$ & $9.778^{\pm .092}$ & $\num{38.488}^{\pm \expon{0.498}}$ \\

50 & $0.372^{\pm .004}$ & $0.554^{\pm .007}$ & $0.659^{\pm .006}$ & $0.295^{\pm .039}$ & $5.263^{\pm .037}$ & $9.803^{\pm .089}$ & $\num{39.190}^{\pm \expon{0.543}}$ \\

60 & $0.373^{\pm .007}$ & $0.554^{\pm .006}$ & $0.664^{\pm .007}$ & $0.289^{\pm .028}$ & \textcolor{blue}{$5.234^{\pm .037}$} & \textcolor{blue}{$9.906^{\pm .072}$} & $\num{39.056}^{\pm \expon{0.504}}$ \\

70 & $0.369^{\pm .007}$ & $0.555^{\pm .006}$ & $0.663^{\pm .006}$ & $0.293^{\pm .032}$ & $5.236^{\pm .025}$ & $9.796^{\pm .080}$ & $\num{39.084}^{\pm \expon{0.541}}$ \\

80 & \textcolor{red}{$0.377^{\pm .005}$} &\textcolor{red}{ $0.558^{\pm .005}$ }& \textcolor{blue}{$0.664^{\pm .006}$} & $0.299^{\pm .032}$ & \textcolor{red}{$5.234^{\pm .021}$} & $9.770^{\pm .077}$ & $\num{39.021}^{\pm \expon{0.641}}$ \\

90 & $0.370^{\pm .004}$ & $0.551^{\pm .005}$ & $0.658^{\pm .005}$ & \textcolor{red}{$0.277^{\pm .029}$} & $5.262^{\pm .030}$ & $9.763^{\pm .067}$ & $\num{38.792}^{\pm \expon{0.615}}$ \\

\midrule
100 & $0.370^{\pm .006}$ & $0.553^{\pm .007}$ & $0.663^{\pm .005}$ & \textcolor{blue}{$0.278^{\pm .030}$} & $5.239^{\pm .027}$ & $9.788^{\pm .103}$ & \textcolor{blue}{$39.52^{\pm 0.7}$} \\
\midrule
Real Motion & $0.511^{\pm .003}$ & $0.703^{\pm .002}$  & $0.796^{\pm .002}$ & $0.001^{\pm .000}$ & $2.972^{\pm .008} $& $9.481^{\pm .107}$ & $47.26^{\pm 0.1}$ \\
\end{tabular}
\end{table*}

\section{Conclusions}
\label{secconclusion}
We introduced FlowMotion, a text-to-motion generation framework leveraging Conditional Flow Matching (CFM). To achieve highly diverse and perceptually consistent 3D character animations with natural motion and low computational costs, FlowMotion incorporates a training objective within CFM that more accurately predicts target motion. We apply this target prediction approach to 3D human motion generation to reduce the inherent jitter associated with CFM while enhancing stability, realism, and computational efficiency. Evaluations on the HumanML3D and KIT datasets demonstrated that FlowMotion achieves a superior balance between fidelity (FID) and smoothness (jitter) while maintaining competitive efficiency against diffusion-based and flow matching baselines. Our framework excels in achieving low jitter values, particularly in the KIT dataset, while also maintaining very competitive FID scores, leading to an excellent Mahalanobis FID-J metric. The sampling process generates motion sequences with the fast generation speeds characteristic of flow matching, a significant improvement over traditional diffusion models, enabling practical deployment in resource-constrained scenarios.

FlowMotion balances computational efficiency and motion quality without relying on oversized architectures. While some methods achieve marginally lower Jitter, our framework prioritizes a holistic trade-off between perceptual realism, temporal consistency, and generation speed, as evidenced by our strong Diversity scores. While our model excels in generating diverse and natural human motions, future research will investigate methods to enhance text-conditional fidelity, specifically targeting improvements in RP Top3, MM-Dist, and MM-Modality metrics. We also plan to extend the framework to applications such as interactive motion editing and full-body articulation, leveraging the model's inherent ability to produce smooth and varied movements.

\section*{Acknowledgments}
This study was financed in part by the Coordenação de Aperfeiçoamento de Pessoal de Nível Superior - Brasil (CAPES) - Finance Code 001. 

\section*{Declaration of Generative AI in Scientific Writing} In the process of preparing this manuscript, the authors employed ChatGPT-4 and Gemini 2.0 to refine the language and enhance the overall readability. After using this tools, the authors meticulously reviewed and revised the content as necessary and assume full responsibility for the final content of this publication.

\bibliographystyle{cag-num-names} 
\bibliography{refs}

\begin{thebibliography}{66}
\providecommand{\natexlab}[1]{#1}
\providecommand{\url}[1]{\texttt{#1}}
\providecommand{\href}[2]{#2}
\providecommand{\path}[1]{#1}
\providecommand{\eprint}[1]{\href{http://arxiv.org/abs/#1}{\path{#1}}}
\providecommand{\DOIprefix}{doi:}
\providecommand{\ArXivprefix}{arXiv:}
\providecommand{\URLprefix}{URL: }
\providecommand{\Pubmedprefix}{pmid:}
\providecommand{\doi}[1]{\href{http://dx.doi.org/#1}{\path{#1}}}
\providecommand{\Pubmed}[1]{\href{pmid:#1}{\path{#1}}}
\providecommand{\BIBand}{and}
\providecommand{\bibinfo}[2]{#2}
\ifx\xfnm\undefined \def\xfnm[#1]{\unskip,\space#1}\fi
%Type = Article
\bibitem[{Jiang et~al.(2024{\natexlab{a}})Jiang, Xie, Li, Yuan, Zhu and
  Zhu}]{jiang2024harmon}
\bibinfo{author}{Jiang\xfnm[ Z]}, \bibinfo{author}{Xie\xfnm[ Y]},
  \bibinfo{author}{Li\xfnm[ J]}, \bibinfo{author}{Yuan\xfnm[ Y]},
  \bibinfo{author}{Zhu\xfnm[ Y]}, \bibinfo{author}{Zhu\xfnm[ Y]}.
\newblock \bibinfo{title}{Harmon: Whole-body motion generation of humanoid
  robots from language descriptions}.
\newblock \bibinfo{journal}{arXiv preprint arXiv:241012773}
  \bibinfo{year}{2024}{\natexlab{a}};.
%Type = Article
\bibitem[{Cheng et~al.(2024)Cheng, Ji, Chen, Yang, Yang and
  Wang}]{cheng2024expressive}
\bibinfo{author}{Cheng\xfnm[ X]}, \bibinfo{author}{Ji\xfnm[ Y]},
  \bibinfo{author}{Chen\xfnm[ J]}, \bibinfo{author}{Yang\xfnm[ R]},
  \bibinfo{author}{Yang\xfnm[ G]}, \bibinfo{author}{Wang\xfnm[ X]}.
\newblock \bibinfo{title}{Expressive whole-body control for humanoid robots}.
\newblock \bibinfo{journal}{arXiv preprint arXiv:240216796}
  \bibinfo{year}{2024};.
%Type = Inproceedings
\bibitem[{Annabi et~al.(2024)Annabi, Ma and Nguyen}]{annabi2024unsupervised}
\bibinfo{author}{Annabi\xfnm[ L]}, \bibinfo{author}{Ma\xfnm[ Z]},
  \bibinfo{author}{Nguyen\xfnm[ SM]}.
\newblock \bibinfo{title}{Unsupervised motion retargeting for human-robot
  imitation}.
\newblock In: \bibinfo{booktitle}{Companion of the 2024 ACM/IEEE International
  Conference on Human-Robot Interaction}. \bibinfo{year}{2024}, p.
  \bibinfo{pages}{204--208}.
%Type = Inproceedings
\bibitem[{Du et~al.(2023)Du, Kips, Pumarola, Starke, Thabet and
  Sanakoyeu}]{du2023avatars}
\bibinfo{author}{Du\xfnm[ Y]}, \bibinfo{author}{Kips\xfnm[ R]},
  \bibinfo{author}{Pumarola\xfnm[ A]}, \bibinfo{author}{Starke\xfnm[ S]},
  \bibinfo{author}{Thabet\xfnm[ A]}, \bibinfo{author}{Sanakoyeu\xfnm[ A]}.
\newblock \bibinfo{title}{Avatars grow legs: Generating smooth human motion
  from sparse tracking inputs with diffusion model}.
\newblock In: \bibinfo{booktitle}{Proceedings of the IEEE/CVF Conference on
  Computer Vision and Pattern Recognition}. \bibinfo{year}{2023}, p.
  \bibinfo{pages}{481--490}.
%Type = Inproceedings
\bibitem[{Luo et~al.(2024)Luo, Cao, Khirodkar, Winkler, Kitani and
  Xu}]{luo2024real}
\bibinfo{author}{Luo\xfnm[ Z]}, \bibinfo{author}{Cao\xfnm[ J]},
  \bibinfo{author}{Khirodkar\xfnm[ R]}, \bibinfo{author}{Winkler\xfnm[ A]},
  \bibinfo{author}{Kitani\xfnm[ K]}, \bibinfo{author}{Xu\xfnm[ W]}.
\newblock \bibinfo{title}{Real-time simulated avatar from head-mounted
  sensors}.
\newblock In: \bibinfo{booktitle}{Proceedings of the IEEE/CVF Conference on
  Computer Vision and Pattern Recognition}. \bibinfo{year}{2024}, p.
  \bibinfo{pages}{571--581}.
%Type = Article
\bibitem[{Yang et~al.(2024)Yang, Li, Wu, Li, Wang, Yu et~al.}]{yang2024smgdiff}
\bibinfo{author}{Yang\xfnm[ H]}, \bibinfo{author}{Li\xfnm[ C]},
  \bibinfo{author}{Wu\xfnm[ Z]}, \bibinfo{author}{Li\xfnm[ G]},
  \bibinfo{author}{Wang\xfnm[ J]}, \bibinfo{author}{Yu\xfnm[ J]}, et~al.
\newblock \bibinfo{title}{Smgdiff: Soccer motion generation using diffusion
  probabilistic models}.
\newblock \bibinfo{journal}{arXiv preprint arXiv:241116216}
  \bibinfo{year}{2024};.
%Type = Article
\bibitem[{Peng et~al.(2022)Peng, Guo, Halper, Levine and Fidler}]{peng2022ase}
\bibinfo{author}{Peng\xfnm[ XB]}, \bibinfo{author}{Guo\xfnm[ Y]},
  \bibinfo{author}{Halper\xfnm[ L]}, \bibinfo{author}{Levine\xfnm[ S]},
  \bibinfo{author}{Fidler\xfnm[ S]}.
\newblock \bibinfo{title}{Ase: Large-scale reusable adversarial skill
  embeddings for physically simulated characters}.
\newblock \bibinfo{journal}{ACM Transactions On Graphics (TOG)}
  \bibinfo{year}{2022};\bibinfo{volume}{41}(\bibinfo{number}{4}):\bibinfo{pages}{1--17}.
%Type = Article
\bibitem[{Starke et~al.(2020)Starke, Zhao, Komura and Zaman}]{starke2020local}
\bibinfo{author}{Starke\xfnm[ S]}, \bibinfo{author}{Zhao\xfnm[ Y]},
  \bibinfo{author}{Komura\xfnm[ T]}, \bibinfo{author}{Zaman\xfnm[ K]}.
\newblock \bibinfo{title}{Local motion phases for learning multi-contact
  character movements}.
\newblock \bibinfo{journal}{ACM Transactions on Graphics (TOG)}
  \bibinfo{year}{2020};\bibinfo{volume}{39}(\bibinfo{number}{4}):\bibinfo{pages}{54--1}.
%Type = Article
\bibitem[{Starke et~al.(2019)Starke, Zhang, Komura and Saito}]{Starke_2019}
\bibinfo{author}{Starke\xfnm[ S]}, \bibinfo{author}{Zhang\xfnm[ H]},
  \bibinfo{author}{Komura\xfnm[ T]}, \bibinfo{author}{Saito\xfnm[ J]}.
\newblock \bibinfo{title}{Neural state machine for character-scene
  interactions}.
\newblock \bibinfo{journal}{ACM Transactions on Graphics}
  \bibinfo{year}{2019};\bibinfo{volume}{38}(\bibinfo{number}{6}):\bibinfo{pages}{1–14}.
\newblock \URLprefix \url{http://dx.doi.org/10.1145/3355089.3356505}.
  \DOIprefix\doi{10.1145/3355089.3356505}.
%Type = Inproceedings
\bibitem[{Jiang et~al.(2024{\natexlab{b}})Jiang, He, Wang, Li, Chen, Huang
  et~al.}]{jiang2024autonomous}
\bibinfo{author}{Jiang\xfnm[ N]}, \bibinfo{author}{He\xfnm[ Z]},
  \bibinfo{author}{Wang\xfnm[ Z]}, \bibinfo{author}{Li\xfnm[ H]},
  \bibinfo{author}{Chen\xfnm[ Y]}, \bibinfo{author}{Huang\xfnm[ S]}, et~al.
\newblock \bibinfo{title}{Autonomous character-scene interaction synthesis from
  text instruction}.
\newblock In: \bibinfo{booktitle}{SIGGRAPH Asia 2024 Conference Papers}.
  \bibinfo{year}{2024}{\natexlab{b}}, p. \bibinfo{pages}{1--11}.
%Type = Misc
\bibitem[{Zhang et~al.(2020)Zhang, Hassan, Neumann, Black and
  Tang}]{zhang2020generating}
\bibinfo{author}{Zhang\xfnm[ Y]}, \bibinfo{author}{Hassan\xfnm[ M]},
  \bibinfo{author}{Neumann\xfnm[ H]}, \bibinfo{author}{Black\xfnm[ MJ]},
  \bibinfo{author}{Tang\xfnm[ S]}.
\newblock \bibinfo{title}{Generating 3d people in scenes without people}.
\newblock \bibinfo{year}{2020}.
\newblock \href{http://arxiv.org/abs/1912.02923}{\tt arXiv:1912.02923}.
%Type = Inproceedings
\bibitem[{Hassan et~al.(2021{\natexlab{a}})Hassan, Ghosh, Tesch, Tzionas and
  Black}]{hassan2021populating}
\bibinfo{author}{Hassan\xfnm[ M]}, \bibinfo{author}{Ghosh\xfnm[ P]},
  \bibinfo{author}{Tesch\xfnm[ J]}, \bibinfo{author}{Tzionas\xfnm[ D]},
  \bibinfo{author}{Black\xfnm[ MJ]}.
\newblock \bibinfo{title}{Populating 3d scenes by learning human-scene
  interaction}.
\newblock In: \bibinfo{booktitle}{Proceedings of the IEEE/CVF Conference on
  Computer Vision and Pattern Recognition}. \bibinfo{year}{2021}{\natexlab{a}},
  p. \bibinfo{pages}{14708--14718}.
%Type = Inproceedings
\bibitem[{Hassan et~al.(2021{\natexlab{b}})Hassan, Ceylan, Villegas, Saito,
  Yang, Zhou et~al.}]{hassan2021stochastic}
\bibinfo{author}{Hassan\xfnm[ M]}, \bibinfo{author}{Ceylan\xfnm[ D]},
  \bibinfo{author}{Villegas\xfnm[ R]}, \bibinfo{author}{Saito\xfnm[ J]},
  \bibinfo{author}{Yang\xfnm[ J]}, \bibinfo{author}{Zhou\xfnm[ Y]}, et~al.
\newblock \bibinfo{title}{Stochastic scene-aware motion prediction}.
\newblock In: \bibinfo{booktitle}{Proceedings of the IEEE/CVF International
  Conference on Computer Vision}. \bibinfo{year}{2021}{\natexlab{b}}, p.
  \bibinfo{pages}{11374--11384}.
%Type = Inproceedings
\bibitem[{Barquero et~al.(2024)Barquero, Escalera and
  Palmero}]{barquero2024seamless}
\bibinfo{author}{Barquero\xfnm[ G]}, \bibinfo{author}{Escalera\xfnm[ S]},
  \bibinfo{author}{Palmero\xfnm[ C]}.
\newblock \bibinfo{title}{Seamless human motion composition with blended
  positional encodings}.
\newblock In: \bibinfo{booktitle}{Proceedings of the IEEE/CVF Conference on
  Computer Vision and Pattern Recognition}. \bibinfo{year}{2024}, p.
  \bibinfo{pages}{457--469}.
%Type = Inproceedings
\bibitem[{Cohan et~al.(2024)Cohan, Tevet, Reda, Peng and van~de
  Panne}]{cohan2024flexible}
\bibinfo{author}{Cohan\xfnm[ S]}, \bibinfo{author}{Tevet\xfnm[ G]},
  \bibinfo{author}{Reda\xfnm[ D]}, \bibinfo{author}{Peng\xfnm[ XB]},
  \bibinfo{author}{van~de Panne\xfnm[ M]}.
\newblock \bibinfo{title}{Flexible motion in-betweening with diffusion models}.
\newblock In: \bibinfo{booktitle}{ACM SIGGRAPH 2024 Conference Papers}.
  \bibinfo{year}{2024}, p. \bibinfo{pages}{1--9}.
%Type = Inproceedings
\bibitem[{Huang et~al.(2023{\natexlab{a}})Huang, Wang, Li, Jia, Liu, Zhu
  et~al.}]{huang2023diffusion}
\bibinfo{author}{Huang\xfnm[ S]}, \bibinfo{author}{Wang\xfnm[ Z]},
  \bibinfo{author}{Li\xfnm[ P]}, \bibinfo{author}{Jia\xfnm[ B]},
  \bibinfo{author}{Liu\xfnm[ T]}, \bibinfo{author}{Zhu\xfnm[ Y]}, et~al.
\newblock \bibinfo{title}{Diffusion-based generation, optimization, and
  planning in 3d scenes}.
\newblock In: \bibinfo{booktitle}{Proceedings of the IEEE/CVF Conference on
  Computer Vision and Pattern Recognition}. \bibinfo{year}{2023}{\natexlab{a}},
  p. \bibinfo{pages}{16750--16761}.
%Type = Article
\bibitem[{Yi et~al.(2024)Yi, Thies, Black, Peng and Rempe}]{yi2024tesmo}
\bibinfo{author}{Yi\xfnm[ H]}, \bibinfo{author}{Thies\xfnm[ J]},
  \bibinfo{author}{Black\xfnm[ MJ]}, \bibinfo{author}{Peng\xfnm[ XB]},
  \bibinfo{author}{Rempe\xfnm[ D]}.
\newblock \bibinfo{title}{Generating human interaction motions in scenes with
  text control}.
\newblock \bibinfo{journal}{arXiv:240410685} \bibinfo{year}{2024};.
%Type = Inproceedings
\bibitem[{Karunratanakul et~al.(2023)Karunratanakul, Preechakul, Suwajanakorn
  and Tang}]{karunratanakul2023guided}
\bibinfo{author}{Karunratanakul\xfnm[ K]}, \bibinfo{author}{Preechakul\xfnm[
  K]}, \bibinfo{author}{Suwajanakorn\xfnm[ S]}, \bibinfo{author}{Tang\xfnm[
  S]}.
\newblock \bibinfo{title}{Guided motion diffusion for controllable human motion
  synthesis}.
\newblock In: \bibinfo{booktitle}{Proceedings of the IEEE/CVF International
  Conference on Computer Vision}. \bibinfo{year}{2023}, p.
  \bibinfo{pages}{2151--2162}.
%Type = Misc
\bibitem[{Petrovich et~al.(2021{\natexlab{a}})Petrovich, Black and
  Varol}]{actorvae}
\bibinfo{author}{Petrovich\xfnm[ M]}, \bibinfo{author}{Black\xfnm[ MJ]},
  \bibinfo{author}{Varol\xfnm[ G]}.
\newblock \bibinfo{title}{Action-conditioned 3d human motion synthesis with
  transformer vae}.
\newblock \bibinfo{year}{2021}{\natexlab{a}}.
\newblock \URLprefix \url{https://arxiv.org/abs/2104.05670}.
  \href{http://arxiv.org/abs/2104.05670}{\tt arXiv:2104.05670}.
%Type = Inbook
\bibitem[{Tevet et~al.(2022)Tevet, Gordon, Hertz, Bermano and
  Cohen-Or}]{Tevet_motionclip2022}
\bibinfo{author}{Tevet\xfnm[ G]}, \bibinfo{author}{Gordon\xfnm[ B]},
  \bibinfo{author}{Hertz\xfnm[ A]}, \bibinfo{author}{Bermano\xfnm[ AH]},
  \bibinfo{author}{Cohen-Or\xfnm[ D]}.
\newblock \bibinfo{title}{MotionCLIP: Exposing Human Motion Generation to CLIP
  Space}.
\newblock \bibinfo{publisher}{Springer Nature Switzerland}.
\newblock ISBN \bibinfo{isbn}{9783031200472}; \bibinfo{year}{2022}, p.
  \bibinfo{pages}{358–374}.
\newblock \URLprefix \url{http://dx.doi.org/10.1007/978-3-031-20047-2_21}.
  \DOIprefix\doi{10.1007/978-3-031-20047-2_21}.
%Type = Article
\bibitem[{Peng et~al.(2017)Peng, Berseth, Yin and Van
  De~Panne}]{peng2017deeploco}
\bibinfo{author}{Peng\xfnm[ XB]}, \bibinfo{author}{Berseth\xfnm[ G]},
  \bibinfo{author}{Yin\xfnm[ K]}, \bibinfo{author}{Van De~Panne\xfnm[ M]}.
\newblock \bibinfo{title}{Deeploco: Dynamic locomotion skills using
  hierarchical deep reinforcement learning}.
\newblock \bibinfo{journal}{Acm transactions on graphics (tog)}
  \bibinfo{year}{2017};\bibinfo{volume}{36}(\bibinfo{number}{4}):\bibinfo{pages}{1--13}.
%Type = Article
\bibitem[{Peng et~al.(2018)Peng, Abbeel, Levine and van~de Panne}]{Peng_2018}
\bibinfo{author}{Peng\xfnm[ XB]}, \bibinfo{author}{Abbeel\xfnm[ P]},
  \bibinfo{author}{Levine\xfnm[ S]}, \bibinfo{author}{van~de Panne\xfnm[ M]}.
\newblock \bibinfo{title}{Deepmimic: example-guided deep reinforcement learning
  of physics-based character skills}.
\newblock \bibinfo{journal}{ACM Transactions on Graphics}
  \bibinfo{year}{2018};\bibinfo{volume}{37}(\bibinfo{number}{4}):\bibinfo{pages}{1–14}.
\newblock \URLprefix \url{http://dx.doi.org/10.1145/3197517.3201311}.
  \DOIprefix\doi{10.1145/3197517.3201311}.
%Type = Inproceedings
\bibitem[{Yuan et~al.(2023)Yuan, Song, Iqbal, Vahdat and
  Kautz}]{yuan2023physdiff}
\bibinfo{author}{Yuan\xfnm[ Y]}, \bibinfo{author}{Song\xfnm[ J]},
  \bibinfo{author}{Iqbal\xfnm[ U]}, \bibinfo{author}{Vahdat\xfnm[ A]},
  \bibinfo{author}{Kautz\xfnm[ J]}.
\newblock \bibinfo{title}{Physdiff: Physics-guided human motion diffusion
  model}.
\newblock In: \bibinfo{booktitle}{Proceedings of the IEEE/CVF International
  Conference on Computer Vision}. \bibinfo{year}{2023}, p.
  \bibinfo{pages}{16010--16021}.
%Type = Inproceedings
\bibitem[{Zhang et~al.(2023{\natexlab{a}})Zhang, Zhang, Cun, Zhang, Zhao, Lu
  et~al.}]{zhang2023generating}
\bibinfo{author}{Zhang\xfnm[ J]}, \bibinfo{author}{Zhang\xfnm[ Y]},
  \bibinfo{author}{Cun\xfnm[ X]}, \bibinfo{author}{Zhang\xfnm[ Y]},
  \bibinfo{author}{Zhao\xfnm[ H]}, \bibinfo{author}{Lu\xfnm[ H]}, et~al.
\newblock \bibinfo{title}{Generating human motion from textual descriptions
  with discrete representations}.
\newblock In: \bibinfo{booktitle}{Proceedings of the IEEE/CVF conference on
  computer vision and pattern recognition}. \bibinfo{year}{2023}{\natexlab{a}},
  p. \bibinfo{pages}{14730--14740}.
%Type = Inproceedings
\bibitem[{Petrovich et~al.(2022)Petrovich, Black and Varol}]{Petrovich2022}
\bibinfo{author}{Petrovich\xfnm[ M]}, \bibinfo{author}{Black\xfnm[ MJ]},
  \bibinfo{author}{Varol\xfnm[ G]}.
\newblock \bibinfo{title}{Temos: Generating diverse human motions from textual
  descriptions}.
\newblock In: \bibinfo{editor}{Avidan\xfnm[ S]}, \bibinfo{editor}{Brostow\xfnm[
  G]}, \bibinfo{editor}{Ciss{\'e}\xfnm[ M]}, \bibinfo{editor}{Farinella\xfnm[
  GM]}, \bibinfo{editor}{Hassner\xfnm[ T]}, editors.
  \bibinfo{booktitle}{Computer Vision -- ECCV 2022}. \bibinfo{address}{Cham}:
  \bibinfo{publisher}{Springer Nature Switzerland}.
\newblock ISBN \bibinfo{isbn}{978-3-031-20047-2}; \bibinfo{year}{2022}, p.
  \bibinfo{pages}{480--497}.
%Type = Inproceedings
\bibitem[{Guo et~al.(2022{\natexlab{a}})Guo, Zou, Zuo, Wang, Ji, Li
  et~al.}]{Guo_2022}
\bibinfo{author}{Guo\xfnm[ C]}, \bibinfo{author}{Zou\xfnm[ S]},
  \bibinfo{author}{Zuo\xfnm[ X]}, \bibinfo{author}{Wang\xfnm[ S]},
  \bibinfo{author}{Ji\xfnm[ W]}, \bibinfo{author}{Li\xfnm[ X]}, et~al.
\newblock \bibinfo{title}{Generating diverse and natural 3d human motions from
  text}.
\newblock In: \bibinfo{booktitle}{2022 IEEE/CVF Conference on Computer Vision
  and Pattern Recognition (CVPR)}. \bibinfo{publisher}{IEEE};
  \bibinfo{year}{2022}{\natexlab{a}},\URLprefix
  \url{http://dx.doi.org/10.1109/cvpr52688.2022.00509}.
  \DOIprefix\doi{10.1109/cvpr52688.2022.00509}.
%Type = Inproceedings
\bibitem[{Hernandez et~al.(2019)Hernandez, Gall and
  Moreno-Noguer}]{hernandez2019human}
\bibinfo{author}{Hernandez\xfnm[ A]}, \bibinfo{author}{Gall\xfnm[ J]},
  \bibinfo{author}{Moreno-Noguer\xfnm[ F]}.
\newblock \bibinfo{title}{Human motion prediction via spatio-temporal
  inpainting}.
\newblock In: \bibinfo{booktitle}{Proceedings of the IEEE/CVF International
  Conference on Computer Vision}. \bibinfo{year}{2019}, p.
  \bibinfo{pages}{7134--7143}.
%Type = Inproceedings
\bibitem[{Degardin et~al.(2022)Degardin, Neves, Lopes, Brito, Yaghoubi and
  Proen{\c{c}}a}]{degardin2022generative}
\bibinfo{author}{Degardin\xfnm[ B]}, \bibinfo{author}{Neves\xfnm[ J]},
  \bibinfo{author}{Lopes\xfnm[ V]}, \bibinfo{author}{Brito\xfnm[ J]},
  \bibinfo{author}{Yaghoubi\xfnm[ E]}, \bibinfo{author}{Proen{\c{c}}a\xfnm[
  H]}.
\newblock \bibinfo{title}{Generative adversarial graph convolutional networks
  for human action synthesis}.
\newblock In: \bibinfo{booktitle}{Proceedings of the IEEE/CVF Winter Conference
  on Applications of Computer Vision}. \bibinfo{year}{2022}, p.
  \bibinfo{pages}{1150--1159}.
%Type = Inproceedings
\bibitem[{Barsoum et~al.(2018)Barsoum, Kender and Liu}]{barsoum2018hp}
\bibinfo{author}{Barsoum\xfnm[ E]}, \bibinfo{author}{Kender\xfnm[ J]},
  \bibinfo{author}{Liu\xfnm[ Z]}.
\newblock \bibinfo{title}{Hp-gan: Probabilistic 3d human motion prediction via
  gan}.
\newblock In: \bibinfo{booktitle}{Proceedings of the IEEE conference on
  computer vision and pattern recognition workshops}. \bibinfo{year}{2018}, p.
  \bibinfo{pages}{1418--1427}.
%Type = Inproceedings
\bibitem[{Ahn et~al.(2018)Ahn, Ha, Choi, Yoo and Oh}]{ahn2018text2action}
\bibinfo{author}{Ahn\xfnm[ H]}, \bibinfo{author}{Ha\xfnm[ T]},
  \bibinfo{author}{Choi\xfnm[ Y]}, \bibinfo{author}{Yoo\xfnm[ H]},
  \bibinfo{author}{Oh\xfnm[ S]}.
\newblock \bibinfo{title}{Text2action: Generative adversarial synthesis from
  language to action}.
\newblock In: \bibinfo{booktitle}{2018 IEEE International Conference on
  Robotics and Automation (ICRA)}. \bibinfo{organization}{IEEE};
  \bibinfo{year}{2018}, p. \bibinfo{pages}{5915--5920}.
%Type = Inproceedings
\bibitem[{Tevet et~al.(2023)Tevet, Raab, Gordon, Shafir, Cohen-or and
  Bermano}]{tevet2022human}
\bibinfo{author}{Tevet\xfnm[ G]}, \bibinfo{author}{Raab\xfnm[ S]},
  \bibinfo{author}{Gordon\xfnm[ B]}, \bibinfo{author}{Shafir\xfnm[ Y]},
  \bibinfo{author}{Cohen-or\xfnm[ D]}, \bibinfo{author}{Bermano\xfnm[ AH]}.
\newblock \bibinfo{title}{Human motion diffusion model}.
\newblock In: \bibinfo{booktitle}{The Eleventh International Conference on
  Learning Representations}. \bibinfo{year}{2023},\URLprefix
  \url{https://openreview.net/forum?id=SJ1kSyO2jwu}.
%Type = Article
\bibitem[{Zhang et~al.(2024)Zhang, Cai, Pan, Hong, Guo, Yang
  et~al.}]{MotionDiffuse}
\bibinfo{author}{Zhang\xfnm[ M]}, \bibinfo{author}{Cai\xfnm[ Z]},
  \bibinfo{author}{Pan\xfnm[ L]}, \bibinfo{author}{Hong\xfnm[ F]},
  \bibinfo{author}{Guo\xfnm[ X]}, \bibinfo{author}{Yang\xfnm[ L]}, et~al.
\newblock \bibinfo{title}{Motiondiffuse: Text-driven human motion generation
  with diffusion model}.
\newblock \bibinfo{journal}{IEEE Transactions on Pattern Analysis and Machine
  Intelligence}
  \bibinfo{year}{2024};\bibinfo{volume}{46}(\bibinfo{number}{6}):\bibinfo{pages}{4115–4128}.
\newblock \URLprefix \url{http://dx.doi.org/10.1109/tpami.2024.3355414}.
  \DOIprefix\doi{10.1109/tpami.2024.3355414}.
%Type = Inproceedings
\bibitem[{Dabral et~al.(2023)Dabral, Mughal, Golyanik and
  Theobalt}]{dabral2023mofusion}
\bibinfo{author}{Dabral\xfnm[ R]}, \bibinfo{author}{Mughal\xfnm[ MH]},
  \bibinfo{author}{Golyanik\xfnm[ V]}, \bibinfo{author}{Theobalt\xfnm[ C]}.
\newblock \bibinfo{title}{Mofusion: A framework for denoising-diffusion-based
  motion synthesis}.
\newblock In: \bibinfo{booktitle}{2023 IEEE/CVF Conference on Computer Vision
  and Pattern Recognition (CVPR)}. \bibinfo{organization}{IEEE};
  \bibinfo{year}{2023}, p. \bibinfo{pages}{9760--9770}.
%Type = Article
\bibitem[{Zhou et~al.(2023)Zhou, Dou, Cao, Liao, Wang, Wang
  et~al.}]{zhou2023emdm}
\bibinfo{author}{Zhou\xfnm[ W]}, \bibinfo{author}{Dou\xfnm[ Z]},
  \bibinfo{author}{Cao\xfnm[ Z]}, \bibinfo{author}{Liao\xfnm[ Z]},
  \bibinfo{author}{Wang\xfnm[ J]}, \bibinfo{author}{Wang\xfnm[ W]}, et~al.
\newblock \bibinfo{title}{Emdm: Efficient motion diffusion model for fast,
  high-quality motion generation}.
\newblock \bibinfo{journal}{arXiv preprint arXiv:231202256}
  \bibinfo{year}{2023};\bibinfo{volume}{2}.
%Type = Article
\bibitem[{Dai et~al.(2024)Dai, Chen, Wang, Liu, Dai and
  Tang}]{dai2024motionlcm}
\bibinfo{author}{Dai\xfnm[ W]}, \bibinfo{author}{Chen\xfnm[ LH]},
  \bibinfo{author}{Wang\xfnm[ J]}, \bibinfo{author}{Liu\xfnm[ J]},
  \bibinfo{author}{Dai\xfnm[ B]}, \bibinfo{author}{Tang\xfnm[ Y]}.
\newblock \bibinfo{title}{Motionlcm: Real-time controllable motion generation
  via latent consistency model}.
\newblock \bibinfo{journal}{arXiv preprint arXiv:240419759}
  \bibinfo{year}{2024};.
%Type = Inproceedings
\bibitem[{Chen et~al.(2023)Chen, Jiang, Liu, Huang, Fu, Chen et~al.}]{mld}
\bibinfo{author}{Chen\xfnm[ X]}, \bibinfo{author}{Jiang\xfnm[ B]},
  \bibinfo{author}{Liu\xfnm[ W]}, \bibinfo{author}{Huang\xfnm[ Z]},
  \bibinfo{author}{Fu\xfnm[ B]}, \bibinfo{author}{Chen\xfnm[ T]}, et~al.
\newblock \bibinfo{title}{Executing your commands via motion diffusion in
  latent space}.
\newblock In: \bibinfo{booktitle}{Proceedings of the IEEE/CVF Conference on
  Computer Vision and Pattern Recognition}. \bibinfo{year}{2023}, p.
  \bibinfo{pages}{18000--18010}.
%Type = Inproceedings
\bibitem[{Lipman et~al.(2023)Lipman, Chen, Ben-Hamu, Nickel and
  Le}]{lipman2022flow}
\bibinfo{author}{Lipman\xfnm[ Y]}, \bibinfo{author}{Chen\xfnm[ RTQ]},
  \bibinfo{author}{Ben-Hamu\xfnm[ H]}, \bibinfo{author}{Nickel\xfnm[ M]},
  \bibinfo{author}{Le\xfnm[ M]}.
\newblock \bibinfo{title}{Flow matching for generative modeling}.
\newblock In: \bibinfo{booktitle}{The Eleventh International Conference on
  Learning Representations}. \bibinfo{year}{2023},\URLprefix
  \url{https://openreview.net/forum?id=PqvMRDCJT9t}.
%Type = Article
\bibitem[{Huang et~al.(2024)Huang, Geng, Luo and Qi}]{huang2024flow}
\bibinfo{author}{Huang\xfnm[ Z]}, \bibinfo{author}{Geng\xfnm[ Z]},
  \bibinfo{author}{Luo\xfnm[ W]}, \bibinfo{author}{Qi\xfnm[ Gj]}.
\newblock \bibinfo{title}{Flow generator matching}.
\newblock \bibinfo{journal}{arXiv preprint arXiv:241019310}
  \bibinfo{year}{2024};.
%Type = Article
\bibitem[{Hu et~al.(2023)Hu, Yin, Ma, Chen, Fernando, Asano
  et~al.}]{hu2023motion}
\bibinfo{author}{Hu\xfnm[ VT]}, \bibinfo{author}{Yin\xfnm[ W]},
  \bibinfo{author}{Ma\xfnm[ P]}, \bibinfo{author}{Chen\xfnm[ Y]},
  \bibinfo{author}{Fernando\xfnm[ B]}, \bibinfo{author}{Asano\xfnm[ YM]},
  et~al.
\newblock \bibinfo{title}{Motion flow matching for human motion synthesis and
  editing}.
\newblock \bibinfo{journal}{arXiv preprint arXiv:231208895}
  \bibinfo{year}{2023};.
%Type = Article
\bibitem[{Lipman et~al.(2024)Lipman, Havasi, Holderrieth, Shaul, Le, Karrer
  et~al.}]{lipman2024flow}
\bibinfo{author}{Lipman\xfnm[ Y]}, \bibinfo{author}{Havasi\xfnm[ M]},
  \bibinfo{author}{Holderrieth\xfnm[ P]}, \bibinfo{author}{Shaul\xfnm[ N]},
  \bibinfo{author}{Le\xfnm[ M]}, \bibinfo{author}{Karrer\xfnm[ B]}, et~al.
\newblock \bibinfo{title}{Flow matching guide and code}.
\newblock \bibinfo{journal}{arXiv preprint arXiv:241206264}
  \bibinfo{year}{2024};.
%Type = Article
\bibitem[{Sutskever(2014)}]{sutskever2014sequence}
\bibinfo{author}{Sutskever\xfnm[ I]}.
\newblock \bibinfo{title}{Sequence to sequence learning with neural networks}.
\newblock \bibinfo{journal}{arXiv preprint arXiv:14093215}
  \bibinfo{year}{2014};.
%Type = Article
\bibitem[{Plappert et~al.(2016)Plappert, Mandery and Asfour}]{kitml}
\bibinfo{author}{Plappert\xfnm[ M]}, \bibinfo{author}{Mandery\xfnm[ C]},
  \bibinfo{author}{Asfour\xfnm[ T]}.
\newblock \bibinfo{title}{The kit motion-language dataset.}
\newblock \bibinfo{journal}{Big Data}
  \bibinfo{year}{2016};\bibinfo{volume}{4}(\bibinfo{number}{4}):\bibinfo{pages}{236--252}.
%Type = Inproceedings
\bibitem[{Petrovich et~al.(2021{\natexlab{b}})Petrovich, Black and
  Varol}]{petrovich2021action}
\bibinfo{author}{Petrovich\xfnm[ M]}, \bibinfo{author}{Black\xfnm[ MJ]},
  \bibinfo{author}{Varol\xfnm[ G]}.
\newblock \bibinfo{title}{Action-conditioned 3d human motion synthesis with
  transformer vae}.
\newblock In: \bibinfo{booktitle}{Proceedings of the IEEE/CVF International
  Conference on Computer Vision}. \bibinfo{year}{2021}{\natexlab{b}}, p.
  \bibinfo{pages}{10985--10995}.
%Type = Inproceedings
\bibitem[{Athanasiou et~al.(2022)Athanasiou, Petrovich, Black and
  Varol}]{athanasiou2022teach}
\bibinfo{author}{Athanasiou\xfnm[ N]}, \bibinfo{author}{Petrovich\xfnm[ M]},
  \bibinfo{author}{Black\xfnm[ MJ]}, \bibinfo{author}{Varol\xfnm[ G]}.
\newblock \bibinfo{title}{Teach: Temporal action composition for 3d humans}.
\newblock In: \bibinfo{booktitle}{2022 International Conference on 3D Vision
  (3DV)}. \bibinfo{organization}{IEEE}; \bibinfo{year}{2022}, p.
  \bibinfo{pages}{414--423}.
%Type = Inproceedings
\bibitem[{Guo et~al.(2022{\natexlab{b}})Guo, Zuo, Wang and Cheng}]{guo2022tm2t}
\bibinfo{author}{Guo\xfnm[ C]}, \bibinfo{author}{Zuo\xfnm[ X]},
  \bibinfo{author}{Wang\xfnm[ S]}, \bibinfo{author}{Cheng\xfnm[ L]}.
\newblock \bibinfo{title}{Tm2t: Stochastic and tokenized modeling for the
  reciprocal generation of 3d human motions and texts}.
\newblock In: \bibinfo{booktitle}{European Conference on Computer Vision}.
  \bibinfo{organization}{Springer}; \bibinfo{year}{2022}{\natexlab{b}}, p.
  \bibinfo{pages}{580--597}.
%Type = Inproceedings
\bibitem[{Radford et~al.(2021)Radford, Kim, Hallacy, Ramesh, Goh, Agarwal
  et~al.}]{clip}
\bibinfo{author}{Radford\xfnm[ A]}, \bibinfo{author}{Kim\xfnm[ JW]},
  \bibinfo{author}{Hallacy\xfnm[ C]}, \bibinfo{author}{Ramesh\xfnm[ A]},
  \bibinfo{author}{Goh\xfnm[ G]}, \bibinfo{author}{Agarwal\xfnm[ S]}, et~al.
\newblock \bibinfo{title}{Learning transferable visual models from natural
  language supervision}.
\newblock In: \bibinfo{booktitle}{International conference on machine
  learning}. \bibinfo{organization}{PMLR}; \bibinfo{year}{2021}, p.
  \bibinfo{pages}{8748--8763}.
%Type = Article
\bibitem[{Jiang et~al.(2023)Jiang, Chen, Liu, Yu, Yu and
  Chen}]{jiang2023motiongpt}
\bibinfo{author}{Jiang\xfnm[ B]}, \bibinfo{author}{Chen\xfnm[ X]},
  \bibinfo{author}{Liu\xfnm[ W]}, \bibinfo{author}{Yu\xfnm[ J]},
  \bibinfo{author}{Yu\xfnm[ G]}, \bibinfo{author}{Chen\xfnm[ T]}.
\newblock \bibinfo{title}{Motiongpt: Human motion as a foreign language}.
\newblock \bibinfo{journal}{Advances in Neural Information Processing Systems}
  \bibinfo{year}{2023};\bibinfo{volume}{36}:\bibinfo{pages}{20067--20079}.
%Type = Article
\bibitem[{Raffel et~al.(2020)Raffel, Shazeer, Roberts, Lee, Narang, Matena
  et~al.}]{raffel2020exploring}
\bibinfo{author}{Raffel\xfnm[ C]}, \bibinfo{author}{Shazeer\xfnm[ N]},
  \bibinfo{author}{Roberts\xfnm[ A]}, \bibinfo{author}{Lee\xfnm[ K]},
  \bibinfo{author}{Narang\xfnm[ S]}, \bibinfo{author}{Matena\xfnm[ M]}, et~al.
\newblock \bibinfo{title}{Exploring the limits of transfer learning with a
  unified text-to-text transformer}.
\newblock \bibinfo{journal}{Journal of machine learning research}
  \bibinfo{year}{2020};\bibinfo{volume}{21}(\bibinfo{number}{140}):\bibinfo{pages}{1--67}.
%Type = Inproceedings
\bibitem[{Bhattacharya et~al.(2021)Bhattacharya, Rewkowski, Banerjee, Guhan,
  Bera and Manocha}]{bhattacharya2021text2gestures}
\bibinfo{author}{Bhattacharya\xfnm[ U]}, \bibinfo{author}{Rewkowski\xfnm[ N]},
  \bibinfo{author}{Banerjee\xfnm[ A]}, \bibinfo{author}{Guhan\xfnm[ P]},
  \bibinfo{author}{Bera\xfnm[ A]}, \bibinfo{author}{Manocha\xfnm[ D]}.
\newblock \bibinfo{title}{Text2gestures: A transformer-based network for
  generating emotive body gestures for virtual agents}.
\newblock In: \bibinfo{booktitle}{2021 IEEE virtual reality and 3D user
  interfaces (VR)}. \bibinfo{organization}{IEEE}; \bibinfo{year}{2021}, p.
  \bibinfo{pages}{1--10}.
%Type = Inproceedings
\bibitem[{Pennington et~al.(2014)Pennington, Socher and
  Manning}]{pennington2014glove}
\bibinfo{author}{Pennington\xfnm[ J]}, \bibinfo{author}{Socher\xfnm[ R]},
  \bibinfo{author}{Manning\xfnm[ CD]}.
\newblock \bibinfo{title}{Glove: Global vectors for word representation}.
\newblock In: \bibinfo{booktitle}{Proceedings of the 2014 conference on
  empirical methods in natural language processing (EMNLP)}.
  \bibinfo{year}{2014}, p. \bibinfo{pages}{1532--1543}.
%Type = Article
\bibitem[{Dhariwal and Nichol(2021)}]{dhariwal2021diffusion}
\bibinfo{author}{Dhariwal\xfnm[ P]}, \bibinfo{author}{Nichol\xfnm[ A]}.
\newblock \bibinfo{title}{Diffusion models beat gans on image synthesis}.
\newblock \bibinfo{journal}{Advances in neural information processing systems}
  \bibinfo{year}{2021};\bibinfo{volume}{34}:\bibinfo{pages}{8780--8794}.
%Type = Misc
\bibitem[{Ramesh et~al.(2022)Ramesh, Dhariwal, Nichol, Chu and
  Chen}]{ramesh2022hierarchical}
\bibinfo{author}{Ramesh\xfnm[ A]}, \bibinfo{author}{Dhariwal\xfnm[ P]},
  \bibinfo{author}{Nichol\xfnm[ A]}, \bibinfo{author}{Chu\xfnm[ C]},
  \bibinfo{author}{Chen\xfnm[ M]}.
\newblock \bibinfo{title}{Hierarchical text-conditional image generation with
  clip latents}.
\newblock \bibinfo{year}{2022}.
\newblock \URLprefix \url{https://arxiv.org/abs/2204.06125}.
  \href{http://arxiv.org/abs/2204.06125}{\tt arXiv:2204.06125}.
%Type = Article
\bibitem[{Huang et~al.(2023{\natexlab{b}})Huang, Park, Wang, Denk, Ly, Chen
  et~al.}]{huang2023noise2music}
\bibinfo{author}{Huang\xfnm[ Q]}, \bibinfo{author}{Park\xfnm[ DS]},
  \bibinfo{author}{Wang\xfnm[ T]}, \bibinfo{author}{Denk\xfnm[ TI]},
  \bibinfo{author}{Ly\xfnm[ A]}, \bibinfo{author}{Chen\xfnm[ N]}, et~al.
\newblock \bibinfo{title}{Noise2music: Text-conditioned music generation with
  diffusion models}.
\newblock \bibinfo{journal}{arXiv preprint arXiv:230203917}
  \bibinfo{year}{2023}{\natexlab{b}};.
%Type = Article
\bibitem[{Zhu et~al.(2023)Zhu, Pang, Chai, Li, Wang, Sun et~al.}]{zhu2023ernie}
\bibinfo{author}{Zhu\xfnm[ P]}, \bibinfo{author}{Pang\xfnm[ C]},
  \bibinfo{author}{Chai\xfnm[ Y]}, \bibinfo{author}{Li\xfnm[ L]},
  \bibinfo{author}{Wang\xfnm[ S]}, \bibinfo{author}{Sun\xfnm[ Y]}, et~al.
\newblock \bibinfo{title}{Ernie-music: Text-to-waveform music generation with
  diffusion models}.
\newblock \bibinfo{journal}{arXiv preprint arXiv:230204456}
  \bibinfo{year}{2023};.
%Type = Inproceedings
\bibitem[{Fei et~al.(2024)Fei, Wu, Ji, Zhang and Chua}]{fei2024dysen}
\bibinfo{author}{Fei\xfnm[ H]}, \bibinfo{author}{Wu\xfnm[ S]},
  \bibinfo{author}{Ji\xfnm[ W]}, \bibinfo{author}{Zhang\xfnm[ H]},
  \bibinfo{author}{Chua\xfnm[ TS]}.
\newblock \bibinfo{title}{Dysen-vdm: Empowering dynamics-aware text-to-video
  diffusion with llms}.
\newblock In: \bibinfo{booktitle}{Proceedings of the IEEE/CVF Conference on
  Computer Vision and Pattern Recognition}. \bibinfo{year}{2024}, p.
  \bibinfo{pages}{7641--7653}.
%Type = Inproceedings
\bibitem[{Kim et~al.(2023)Kim, Kim and Choi}]{kim2023flame}
\bibinfo{author}{Kim\xfnm[ J]}, \bibinfo{author}{Kim\xfnm[ J]},
  \bibinfo{author}{Choi\xfnm[ S]}.
\newblock \bibinfo{title}{Flame: Free-form language-based motion synthesis \&
  editing}.
\newblock In: \bibinfo{booktitle}{Proceedings of the AAAI Conference on
  Artificial Intelligence}; vol.~\bibinfo{volume}{37}. \bibinfo{year}{2023}, p.
  \bibinfo{pages}{8255--8263}.
%Type = Article
\bibitem[{Liu(2019)}]{liu2019roberta}
\bibinfo{author}{Liu\xfnm[ Y]}.
\newblock \bibinfo{title}{Roberta: A robustly optimized bert pretraining
  approach}.
\newblock \bibinfo{journal}{arXiv preprint arXiv:190711692}
  \bibinfo{year}{2019};\bibinfo{volume}{364}.
%Type = Inproceedings
\bibitem[{Zhang et~al.(2023{\natexlab{b}})Zhang, Guo, Pan, Cai, Hong, Li
  et~al.}]{zhang2023remodiffuse}
\bibinfo{author}{Zhang\xfnm[ M]}, \bibinfo{author}{Guo\xfnm[ X]},
  \bibinfo{author}{Pan\xfnm[ L]}, \bibinfo{author}{Cai\xfnm[ Z]},
  \bibinfo{author}{Hong\xfnm[ F]}, \bibinfo{author}{Li\xfnm[ H]}, et~al.
\newblock \bibinfo{title}{Remodiffuse: Retrieval-augmented motion diffusion
  model}.
\newblock In: \bibinfo{booktitle}{Proceedings of the IEEE/CVF International
  Conference on Computer Vision}. \bibinfo{year}{2023}{\natexlab{b}}, p.
  \bibinfo{pages}{364--373}.
%Type = Inproceedings
\bibitem[{Rombach et~al.(2022)Rombach, Blattmann, Lorenz, Esser and
  Ommer}]{rombach2022high}
\bibinfo{author}{Rombach\xfnm[ R]}, \bibinfo{author}{Blattmann\xfnm[ A]},
  \bibinfo{author}{Lorenz\xfnm[ D]}, \bibinfo{author}{Esser\xfnm[ P]},
  \bibinfo{author}{Ommer\xfnm[ B]}.
\newblock \bibinfo{title}{High-resolution image synthesis with latent diffusion
  models}.
\newblock In: \bibinfo{booktitle}{Proceedings of the IEEE/CVF conference on
  computer vision and pattern recognition}. \bibinfo{year}{2022}, p.
  \bibinfo{pages}{10684--10695}.
%Type = Article
\bibitem[{Chen et~al.(2018)Chen, Rubanova, Bettencourt and Duvenaud}]{chennn}
\bibinfo{author}{Chen\xfnm[ TQ]}, \bibinfo{author}{Rubanova\xfnm[ Y]},
  \bibinfo{author}{Bettencourt\xfnm[ J]}, \bibinfo{author}{Duvenaud\xfnm[ D]}.
\newblock \bibinfo{title}{Neural ordinary differential equations}.
\newblock \bibinfo{journal}{CoRR}
  \bibinfo{year}{2018};\bibinfo{volume}{abs/1806.07366}.
\newblock \URLprefix \url{http://arxiv.org/abs/1806.07366}.
  \href{http://arxiv.org/abs/1806.07366}{\tt arXiv:1806.07366}.
%Type = Misc
\bibitem[{Adewole et~al.(2024)Adewole, Giwa, Nerrise, Osifeko and
  Oyedeji}]{adewole2024humanmotionsynthesisdiffusion}
\bibinfo{author}{Adewole\xfnm[ M]}, \bibinfo{author}{Giwa\xfnm[ O]},
  \bibinfo{author}{Nerrise\xfnm[ F]}, \bibinfo{author}{Osifeko\xfnm[ M]},
  \bibinfo{author}{Oyedeji\xfnm[ A]}.
\newblock \bibinfo{title}{Human motion synthesis - a diffusion approach for
  motion stitching and in-betweening}.
\newblock \bibinfo{year}{2024}.
\newblock \URLprefix \url{https://arxiv.org/abs/2409.06791}.
  \href{http://arxiv.org/abs/2409.06791}{\tt arXiv:2409.06791}.
%Type = Article
\bibitem[{Duan et~al.(2021)Duan, Shi, Zou, Lin, Qian, Zhang
  et~al.}]{singleshot}
\bibinfo{author}{Duan\xfnm[ Y]}, \bibinfo{author}{Shi\xfnm[ T]},
  \bibinfo{author}{Zou\xfnm[ Z]}, \bibinfo{author}{Lin\xfnm[ Y]},
  \bibinfo{author}{Qian\xfnm[ Z]}, \bibinfo{author}{Zhang\xfnm[ B]}, et~al.
\newblock \bibinfo{title}{Single-shot motion completion with transformer}.
\newblock \bibinfo{journal}{CoRR}
  \bibinfo{year}{2021};\bibinfo{volume}{abs/2103.00776}.
\newblock \URLprefix \url{https://arxiv.org/abs/2103.00776}.
  \href{http://arxiv.org/abs/2103.00776}{\tt arXiv:2103.00776}.
%Type = Article
\bibitem[{Loshchilov(2017)}]{loshchilov2017decoupled}
\bibinfo{author}{Loshchilov\xfnm[ I]}.
\newblock \bibinfo{title}{Decoupled weight decay regularization}.
\newblock \bibinfo{journal}{arXiv preprint arXiv:171105101}
  \bibinfo{year}{2017};.
%Type = Inproceedings
\bibitem[{Guo et~al.(2020)Guo, Zuo, Wang, Zou, Sun, Deng
  et~al.}]{guo2020action2motion}
\bibinfo{author}{Guo\xfnm[ C]}, \bibinfo{author}{Zuo\xfnm[ X]},
  \bibinfo{author}{Wang\xfnm[ S]}, \bibinfo{author}{Zou\xfnm[ S]},
  \bibinfo{author}{Sun\xfnm[ Q]}, \bibinfo{author}{Deng\xfnm[ A]}, et~al.
\newblock \bibinfo{title}{Action2motion: Conditioned generation of 3d human
  motions}.
\newblock In: \bibinfo{booktitle}{Proceedings of the 28th ACM International
  Conference on Multimedia}. \bibinfo{year}{2020}, p.
  \bibinfo{pages}{2021--2029}.
%Type = Inproceedings
\bibitem[{Mahmood et~al.(2019)Mahmood, Ghorbani, Troje, Pons-Moll and
  Black}]{mahmood2019amass}
\bibinfo{author}{Mahmood\xfnm[ N]}, \bibinfo{author}{Ghorbani\xfnm[ N]},
  \bibinfo{author}{Troje\xfnm[ NF]}, \bibinfo{author}{Pons-Moll\xfnm[ G]},
  \bibinfo{author}{Black\xfnm[ MJ]}.
\newblock \bibinfo{title}{Amass: Archive of motion capture as surface shapes}.
\newblock In: \bibinfo{booktitle}{Proceedings of the IEEE/CVF international
  conference on computer vision}. \bibinfo{year}{2019}, p.
  \bibinfo{pages}{5442--5451}.
%Type = Inproceedings
\bibitem[{Vaswani et~al.(2017)Vaswani, Shazeer, Parmar, Uszkoreit, Jones, Gomez
  et~al.}]{allattention}
\bibinfo{author}{Vaswani\xfnm[ A]}, \bibinfo{author}{Shazeer\xfnm[ N]},
  \bibinfo{author}{Parmar\xfnm[ N]}, \bibinfo{author}{Uszkoreit\xfnm[ J]},
  \bibinfo{author}{Jones\xfnm[ L]}, \bibinfo{author}{Gomez\xfnm[ AN]}, et~al.
\newblock \bibinfo{title}{Attention is all you need}.
\newblock In: \bibinfo{editor}{Guyon\xfnm[ I]}, \bibinfo{editor}{Luxburg\xfnm[
  UV]}, \bibinfo{editor}{Bengio\xfnm[ S]}, \bibinfo{editor}{Wallach\xfnm[ H]},
  \bibinfo{editor}{Fergus\xfnm[ R]}, \bibinfo{editor}{Vishwanathan\xfnm[ S]},
  et~al., editors. \bibinfo{booktitle}{Advances in Neural Information
  Processing Systems}; vol.~\bibinfo{volume}{30}. \bibinfo{publisher}{Curran
  Associates, Inc.}; \bibinfo{year}{2017},\URLprefix
  \url{https://proceedings.neurips.cc/paper_files/paper/2017/file/3f5ee243547dee91fbd053c1c4a845aa-Paper.pdf}.

\end{thebibliography}

\end{document}